\documentstyle[prb,preprint,aps]{revtex}

\begin{document}

\title{
He Scattering from Compact Clusters and from
Diffusion-Limited Aggregates on
Surfaces:\\
Observable Signatures of Structure
}

\author{D.A. Hamburger$^{a,b}$, A.T. Yinnon$^{b}$, I.
Farbman$^{b}$, A. Ben-Shaul$^{b}$ and R.B. Gerber$^{b,c}$}
\address{
$^a$Department of Physics, The Hebrew University of
Jerusalem, Jerusalem
91904, Israel\\
$^b$Department of Physical Chemistry and The Friz Haber
Center for Molecular
Dynamics, The Hebrew University of Jerusalem, Jerusalem
91904, Israel\\
$^c$Department of Chemistry, University of California -
Irvine, Irvine, CA
92717, USA
}

\maketitle

\begin{abstract}
\newline{}
The angular intensity distribution of He beams scattered
from compact clusters
and from diffusion limited aggregates, epitaxially grown on
metal surfaces, is
investigated theoretically. The purpose is twofold: to
distinguish compact
cluster structures from diffusion limited aggregates, and
to find observable
{\em signatures} that can characterize the compact clusters
at the atomic
level of detail. To simplify the collision dynamics, the
study is carried out
in the framework of
the sudden approximation, which assumes that momentum
changes perpendicular to
the surface are large compared with momentum transfer due
to surface
corrugation. The diffusion limited aggregates on which the
scattering
calculations were done, were generated by kinetic Monte
Carlo simulations. It
is demonstrated, by focusing on the example of compact Pt
Heptamers, that
signatures of structure of compact clusters may indeed be
extracted from the
scattering distribution. These signatures enable both an
experimental
distinction between diffusion limited aggregates and
compact clusters, and a
determination of the cluster structure. The characteristics
comprising the
signatures are, to varying degrees, the Rainbow,
Fraunhofer, specular and constructive interference peaks,
all seen in the
intensity
distribution. It is also shown, how the distribution of
adsorbate heights
above the metal surface can be obtained by an analysis of
the specular peak
attenuation. The results contribute to establishing He
scattering as a
powerful tool in the investigation of surface disorder and
epitaxial growth on
surfaces, alongside with STM.
\end{abstract}

\newpage

\section{Introduction}
\label{intro}

The process of epitaxial growth of metal or semiconductor
films on a surface
presents some major theoretical, experimental and
technological challenges. In
epitaxial growth, metal or semiconductor atoms are adsorbed
on a corresponding
surface under thermal conditions, to form two- and
three-dimensional
structures on top of it. The physical and chemical
properties are determined
by the final form of these structures.  These may be of
dramatic importance,
e.g, in the production of electronic devices.  One of the
most exciting
aspects of epitaxial growth kinetics, is that it {\bf
prepares disordered
structures} in the intermediate stages. The disorder
manifests itself in the
formation of various types of clusters or
diffusion-limited-aggregates (DLA)
on top of the surface.  These structures may be monolayers
(usually at high
temperatures, when the diffusivity is large), in which case
the disorder is
two-dimensional, or they may be composed of several layers,
giving rise to
disorder in three dimensions. Epitaxially grown structures
of this type have
for some time now received wide attention, as they offer an
exceptional
opportunity for both experimental and theoretical study of
disorder. Clusters
in particular play an important role in the
epitaxial-growth kinetics, as it
is the continual process of their formation, restructuring
and dissociation,
through which growth actually takes place.  Issues such as
the structure,
stability and surface diffusion of metal clusters have been
addressed
experimentally by Wang and Ehrlich \cite{Ehrlich:Ir}.
Rosenfeld et al.
\cite{Comsa:heptamers} have recently suggested the
intriguing possibility of
the formation over a certain temperature and coverage
range, of a high density
of small, stable and compact clusters of seven ad-atoms of
Pt (Heptamers),
during the process of epitaxial growth of Pt on Pt(111). A
recent theoretical
first-principles study by Stumpf and Scheffler
\cite{Stumpf} predicts
hexagonally shaped islands for Al on Al(111). These
developments tentatively
suggest that metastable states of systems consisting of
clusters of
well-defined shape, and perhaps size may exist. Regarding
DLA, several
authors
\cite{DePristo:DLA1,DePristo:DLA2,Metiu:DLA1,Metiu:DLA2}
have
presented simulations which reveal the complexity
anticipated in the shapes of
these structures, in which no regular clusters can be
discerned.  As these and
other studies show, a large variety of structures
contribute to surface
disorder. No satisfactory and comprehensive theory of the
epitaxial growth
process is as of yet available, much due to the absence of
reliable
interaction potentials between the ad-atoms and the
surface.  Progress at this
stage thus hinges critically on experiments. To a large
extent the information
obtained from different modern methods is complementary.
 Since the various
techniques probe different aspects and distance scales of
the surface
structures, the most comprehensive picture will probably be
attained via an
integration of results from all techniques. TEAS and in
particular He-beams,
has already proved among the most useful tools in analyzing
surface phenomena.
Experiments by Poelsema and co-workers
\cite{Comsa:defects1,Comsa:defects2,Comsa:defects3,Comsa:defects4}
and Ernst
and coworkers \cite{Ernst:1,Ernst:2,Ernst:3,Ernst:4} have
employed
He-scattering in dealing with questions concerned with
scattering from
isolated surface defects, and were so able to extract a
wealth of information
on the defects and their mutual attractions. The
attenuation of specular
intensities, in particular, proves an extremely useful tool
in these
investigations.  Theoretical studies by Gerber and
co-workers
\cite{Benny:RBs,Benny:1,Benny:2,Benny:3} utilizing both
time independent
(Sudden approximation) and time-dependent methods (quantum
wave-packets), have
analyzed the calculated angular distribution and specular
peak attenuation.
These authors have been able to identify connections
between the features of
the distribution and the attenuation, and microscopic
characteristics related
to the surface defects.  An important advantage offered by
He-scattering, is
that He is capable of probing structure both at the {\bf
local} and {\bf
global} levels. The interpretation of He-scattering
experiments, however, is
rather more involved than that of the direct-imaging
techniques such as FIM
and STM.  The purpose of the present article is to address
the questions of
whether ``signatures'' of the various shapes assumed by the
clusters and DLA
are present in the spectra obtained with He-scattering
experiments, and to
find how they may be extracted with the proper analysis.
The answers to these
questions are crucial to the understanding of
epitaxial-growth induced surface
disorder, and to enhance the usefulness of TEAS as an
indispensable tool in
the study of such surface phenomena.  The experiments by
Rosenfeld et al.
\cite{Comsa:heptamers} and Bott et al. \cite{Comsa:STM1} on
Pt and by Ehrlich
et al. \cite{Ehrlich:Ir} on Ir suggest that under proper
conditions compact
clusters may be observed as a dominant phase, whereas DLA
will predominate at
other conditions \cite{Comsa:STM1}. This is supported by
Stumpf and
Scheffler's theoretical calculations for Al \cite{Stumpf}.
It will be our main
purpose in this article to demonstrate that compact
clusters can easily be
distinguished from DLA in a He-scattering experiment.
 Furthermore we hope to
demonstrate, using the example of compact Heptamers as a
case study for
clusters, that specific structure-related characteristics
of compact clusters
can indeed be extracted from an analysis of the angular
intensity
distribution, and that these characteristics, when
combined, provide an almost
unique signature of the structure. Thus by a detailed
examination of the
angular intensity distribution calculated from several
model systems, we seek
to show that surface-adsorbed structures exhibiting various
types of disorder,
can be {\bf characterized} by means of He-scattering
experiments. The
structure of the article is as follows: In
Section~\ref{systems} we discuss
our model systems.  Section~\ref{methods} briefly discusses
the methods
employed for studying the scattering intensities.
Section~\ref{results}
describes the results of scattering calculations from our
model systems.
Concluding remarks are brought in
section~\ref{conclusions}.

\section{The Model Systems}
\label{systems}

The present study deals with the scattering of a He beam
from various types of
disordered structures, considered pertinent to epitaxial
growth modes. The
possible formation of such structures in kinetic
simulations of growth is also
treated briefly. We shall treat two types of structures:
\begin{enumerate}
\item{
{\em Compact Clusters on a Surface.} \\
Within this family, disorder will be represented in an
increasing degree of
complexity, by considering three related models:
}
\begin{enumerate}
\item{
A single Heptamer on an otherwise completely flat Pt(111)
surface.
}
\item{
A collection of Heptamers on the same Pt surface, adsorbed
at random
positions, but having orientations compatible with the
underlying hexagonal
symmetry.
}
\item{
A collection of Heptamers on the same Pt surface, adsorbed
both at random
positions and with random orientations in the surface
plane.
}
\end{enumerate}
\item{
{\em Diffusion Limited Aggregates on a Surface.} \\
Here we considered non-compact structures, grown under
kinetic conditions.
}
\end{enumerate}

The clusters and the underlying surface were treated as a
static, nonvibrating
target in the scattering calculations. Clearly, vibrations
of the adsorbate
atoms and the surface are expected to affect quantitatively
the scattering
distributions measured in TEAS experiments.  However,
results of calculations
using a rigid, nonvibrating surface system should be useful
at least for
studying the main qualitative effects.  Moreover, for
diffraction scattering
from crystalline surfaces, the effect of surface vibrations
on the scattering
intensities can be represented approximately by a simple
Debye-Waller factor
\cite{Benny:DW}. A similar description is expected to be
successful also for
the scattering intensities from clusters on surfaces. In
the context of
studies of the specular attenuation, work by Poelsema,
Comsa and coworkers
\cite{Comsa:book}
has demonstrated that treating the surface as static is a
very useful
approximation in analyzing specular scattering from
disordered surfaces.  The
Pt surface in the scattering calculations reported here was
treated as
completely flat. For a very low corrugation surface such as
Pt(111), this
should be a good approximation.  We now describe in some
detail the three
models of compact Heptamer clusters and the model of
kinetic growth of DLA on
the surface, all of which were used for the scattering
calculations.

\subsection{Single Heptamer on a Flat Pt Surface}
\label{systems-single}

In this section we describe a model for a single, symmetric
and static
Heptamer on a smooth metal surface. The parameters were
chosen to represent a
Pt Heptamer on a Pt(111) surface. Although we do not expect
a TEAS experiment
to ever probe a surface with just a single Heptamer
adsorbed upon it, the
study of the angular intensity distribution produced by
such an arrangement is
crucial in understanding the specific Heptamer related
features. As a
motivation for our choice of potential, consider the role
of the conduction
electrons.  Clearly the Heptamer is immersed in the sea of
surface conduction
electrons.  However, we expect the surface to be much
smoother than the
Heptamer's top, due to a flow of electrons from the top to
the surface. The
extent to which the Heptamer's nuclei lie bare determines
both the Heptamer's
height above the surface and its top's corrugation. These
are very important
parameters in determining the scattering features. A
simple, albeit inexact,
class of potentials which satisfy the requirement of a
corrugated
Heptamer-top, are those which are pairwise additive in the
top.  We assume the
Heptamer to be composed of an isolated cluster of seven Pt
atoms adsorbed on a
flat Pt surface, with six Pt atoms forming a perfect
hexagon and the seventh
atom located at its center. The interaction between the gas
phase He and the
Pt surface with its adsorbed cluster is assumed additive
and is represented by
the following potential function:

\begin{equation}
{V}({\bf r}) = {V_s}(z) + \sum_{i =
1}^{7}{V_{He-Pt}}(\left|{\bf r} - {\bf
r_i}\right|) \label{eq:singleHep}
\end{equation}

where ${\bf r}=(x,y,z)$, ${\rm V_s}(z)$ represents the
interaction between He
and a flat Pt(111) surface, and ${\rm V_{\em
He-Pt}}(\left|{\bf r} - {\bf
r_i}\right|)$ represents the interaction between He and the
$i_{th}$ Pt atom
in the cluster, located at ${\bf r_i}$. The coordinate z
measures the distance
of the He atom from the surface plane, while $(x,y)$ are
the coordinates in
parallel to the surface plane. The functional form of ${\rm
V_s}(z)$ was
chosen as:

\begin{equation}
{V_s}(z) = {D_s}\,\left({{{{{z_m}}^9}}\over {{z^9}}} -
{{3\,{{{z_m}}^3}}\over
{{z^3}}}\right) \label{eq:Vs}
\end{equation}

which is a surface scattering version of a Lennard-Jones
6-12 potential. The
value $D_s=4.0 meV (=8.464 \cdot 10^{-5} a.u.)$, given by
Harris et al.
\cite{Harris:RHEED} was employed, and $z_m$ was taken as
$7.87 a.u.$, which
gives a steepness typical of atom-atom interactions with
metal surfaces. For
$V_{He-Pt}$ a Lennard-Jones $6-12$ potential was used:

\begin{equation}
{V_{He-Pt}}(r) = {{{c_{12}}}\over {{r^{12}}}} -
{{{c_6}}\over {{r^6}}}
\label{eq:VHePt}
\end{equation}

We assumed that the $c_6$ interaction parameter between He
and a heavy atom X
scales linearly with the polarizability of X. The
polarizability, in turn, was
estimated from the available atomic volume. In our
estimate, the $c_6$
parameter so obtained is sufficiently reliable for the
qualitative purposes of
this paper. Thus:\\
\indent
$c_6$(Pt-He) $\sim c_6$(Xe-He) Vol(Pt)/Vol(Xe) = 6.322
a.u.\\
The $c_{12}$ parameter was extracted from eq.(\ref{eq:Vs})
by the formula:

\begin{equation}
{V_s}(z) = {1 \over
    {{S^{{3\over 2}}}}} \, {2\,\pi \,\left( \int
_{z}^{\infty}
         dr\,\left( {r^2} - r\,z \right) \,{V_{He-Pt}}(r)
\right) }
\end{equation}

where $S$ is the unit-cell area. This equation is based on
taking the
continuum limit of a sum of pairwise He/Pt interactions. It
gives a value of
$c_{12}=3.251 \cdot 10^7$a.u. We do not have ab-initio
calculations available
to provide reliable estimates
of our additive-potentials model, and thus we do not expect
the interaction
potential eq.(\ref{eq:singleHep}) to be quantitatively
correct, but we expect
this
potential to be realistic on a semiquantitative footing.
This should suffice
for the purpose of searching for qualitative effects in the
angular
distribution.

\subsection{Randomly Adsorbed Heptamers}
\label{systems-random}

The situation we expect to find in a real experiment is a
{\em distribution}
of clusters of various sizes and shapes: isolated Heptamers
are probably not
realized at the coverages of interest for epitaxial growth.
Consider as a
conjecture the existence of Heptamer cluster phases. If
confirmed after further study, this distribution will
exhibit a pronounced
peak for symmetrical Heptamers.  Hence a more realistic
treatment of the
scattering off Heptamers must involve a distribution of
Heptamers on a
surface. The Heptameric-phase conjecture does, however,
leave us with a
freedom to consider two models of a distribution of
adsorbed Heptamers.  These
correspond to different degrees of disorder, and follow
from the possibility
of either a high or low barrier for rotation of a
Heptamer-cluster as a whole,
as explained below. In this section we investigate the
scattering from the two
corresponding distributions of Heptamers, while ignoring
the possible
existence of other cluster types. We expect the other
cluster types not to
alter the central features of the intensity spectrum,
provided the
distribution is peaked sharply enough about the Heptamers.
In this sense the
present framework is an adequate tool for our purpose of
providing signatures
of Heptamers. In the two models we investigated in this
section we considered
the case of randomly adsorbed Heptamers. In both, the
center of each Heptamer
was chosen randomly from a uniform distribution, and then
placed on the
nearest lattice site. We excluded those Heptamers that
overlapped with or
bordered on other Heptamers, i.e, those where the distance
of Pt atoms
belonging to two different clusters was less than two unit
cells apart. We
distinguish between two possibilities:

\subsubsection{Translationally Random Heptamers on Pt(111)}
\label{systems-random1}

Since it is plausible that the orientation of the Heptamers
is determined by
the surface structure, we located each cluster Pt atom at
the center of the
triangles that make up the underlying hexagonal structure
of the surface Pt
atoms (i.e, a lattice site).  This corresponds to a high
barrier for rotation
of a Heptamer-cluster as a whole. It leads to a 6-fold
symmetry in the
orientations of the Heptamers. Part of the surface is shown
in
figure~\ref{fig:oriented}.

\subsubsection{Translationally and Rotationally Random
Heptamers on Pt(111)}
\label{systems-random2}

Though the surface-induced orientation described
in~\ref{systems-random1}
seems intuitively physically correct, we cannot exclude a
completely randomly
distributed orientation of the Heptamers, corresponding to
a low barrier for
rotation of a Heptamer-cluster as a whole.  Therefore we
investigated a second
model, where also the orientation of the Heptamers is
chosen randomly from a
uniform distribution, as can be seen in
figure~\ref{fig:random}.  For both
cases the coverage of the adsorbed Pt atoms was 5\% of a
monolayer.  The He /
Pt cluster+surface interaction for these models is given by
eq.(\ref{eq:singleHep}), with the exception of the second
term which now
contains an additional sum over all N Heptamers:

\begin{equation}
{V}({\bf r}) = \sum_{n = 1}^{{N}} \sum_{i = 1}^{7}{
V_{He-Pt}}(|{\bf r} - {{\bf r}_{i,n}}|) + {Vs}(z)
\end{equation}

Here ${{\bf r}_{i,n}} = (x_{i,n},y_{i,n},z_{i,n})$
represents the location of
the $i_{th}$ Pt atom in the $n_{th}$ Heptamer.

\subsection{Diffusion Limited Aggregates Grown by Kinetic
Monte Carlo (KMC)}
\label{systems-DLA}

We now proceed to consider not only compact clusters, but
structures that can
develop in kinetic conditions. The main factors determining
the shapes of the
resulting structures are:
\begin{enumerate}
\item{
the deposition rate of the Pt atoms
}
\item{
spatial migration of the Pt atoms upon the surface
}
\end{enumerate}
There is an extensive literature dealing with the modeling
of the process of
diffusion and aggregation upon a homogeneous surface
\cite{Kang1}. The
approach that will be taken here to simulate the
time-dependent structures
obtained in the course of deposition and migration, is the
KMC scheme
\cite{Kang2}. The essence of the approach is assigning
probabilities for the
hopping of adsorbed atoms between allowed sites upon the
surface, which is
assumed a lattice. To obtain probabilities for hops between
the different
configurations, assumptions about energetics and
interactions must be made.
We shall present a kinetic model for growth of monolayer Pt
islands on a flat
Pt surface. Our purpose in examining this model was to
provide realistic
adsorbate structures for the He-scattering calculations.
 This enabled us to
investigate possible ways of discriminating between these
structures and
formations of Heptamers. In order to produce realistic
arrangements of
adatoms on the surface we performed Kinetic Monte Carlo
(KMC) simulations,
using a model of pairwise additive interactions and a Time
Dependent Monte
Carlo (TDMC) procedure.  The simulation took place on a
hexagonal lattice of
$100\cdot 100$ unit cells with periodic boundary
conditions.  The simulation
starts with two particles on the surface diffusing until a
third particle
adsorbs. This period is typically 0.05 sec (for simulating
an adsorption rate
of $500 sec/ML$ \cite{Comsa:heptamers}.)  corresponding to
about $10^6$ MC
steps at room temperature. Then the three particles
diffuse, a fourth one
adsorbs, and so on until 500 particles, corresponding to
5\% coverage, diffuse
on the surface for 0.05 sec. The configurations attained
are not at
equilibrium.

\paragraph{The Interactions Model}
The simulations were performed using two different kinetic
models for lateral
interactions between the adsorbed particles. In the first
model, every
particle was assigned an energy according to the number of
its nearest
neighbors

\begin{equation}
E(n) = E_0 + n\, \epsilon \label{eq:w-const}
\end{equation}

where $E_0$ is the activation energy for diffusion at zero
coverage, $n$ is
the number of nearest neighbors and $\epsilon$ is the
nearest neighbor
interaction energy. Particle diffusion takes place via
random walks between
nearest neighbor sites. The hopping rate (transition
probability per unit
time) of a particle from site $i$, where it has $n_i$
nearest neighbors, to a
neighboring site with $n_f$ nearest neighbors, is taken as

\begin{equation}
\omega_{i \rightarrow f}=\nu \, e^{{-E(n_i)} \over {k\, T}}
= \omega_0 \,
e^{{-n_i
\, \epsilon} \over {k\, T}} \label{eq:w-if}
\end{equation}

where $\omega_0 = \nu \, exp[-E_0/(k\, T)]$ is the hopping
frequency of an
``isolated'' particle. $\tau_0=1/ \omega_0$, The average
time interval between
successive moves on the bare (zero coverage) surface, sets
the unit of time
for simulations performed at a given temperature $T$.
Clearly, $\omega_{i
\rightarrow f}$ satisfies detailed balance, since
$\omega_{i
\rightarrow f}/\omega_{f \rightarrow i} =
exp[-(E_f-E_i)/(k\, T)]$ with
$E_i=E(n_i)$ and $E_f=E(n_f)$. Note that according to
eq.(\ref{eq:w-if}),
the transition rate depends only on the initial state. A
possible dynamical
interpretation of this model, originally suggested by
Uebing and Gomer
\cite{Uebing}, is that the rate of the transition from
state $i$ to state $f$,
is governed by the rate at which the particle in state $i$
escapes into a
transition state (saddle point) which is the same for all
possible transitions
into different final states. Examples of structures
produced by this model are
shown in figure \ref{fig:islands}.\\
In the second model we continued to use eq.(\ref{eq:w-if})
for the transition
rates, but now instead of eq.(\ref{eq:w-const}), the energy
of a particle in a
given configuration is:

\begin{equation}
E(n,s) = E_0 + n\, \epsilon (s) \label{eq:w-s}
\end{equation}

with $n$ denoting, as before, the number of nearest
neighbors of the particle
under consideration, and $s$ being the size of the cluster
the particle
belongs to. According to this model, the nearest neighbor
interaction energy
$\epsilon (s)$, depends on the cluster size (thus,
indirectly, taking into
account the effects of non-additive interactions). The
assumption on the
nature of the interaction is critical. It is clear that by
assuming certain
types of dependence of the microcluster energy upon shape
and size, one can
ultimately obtain any chosen type of kinetic behavior and
global structure. We
believe that the dependences we used, while definitely not
quantitatively
valid, represent at least qualitative a plausible energetic
behavior of a
system of this type. The values of $\epsilon (s)$ (for Pt
atoms on Pt(111))
were taken from the work of Rosenfeld et al.
\cite{Comsa:heptamers1}, who
used eq.(\ref{eq:w-s}) in their rate equation study of Pt
diffusion and
aggregation.  More specifically, these authors used: \\
$\epsilon (s)= \epsilon (2)-(s-2) \, \Delta \epsilon$, with
$\epsilon
(2)=0.300 eV$, and $\Delta \epsilon=0.015 eV$ for $2 \leq s
\leq 6$, and: \\
$\epsilon (s)= \epsilon (6)-(s-6) \, \Delta \epsilon$, with
$\epsilon
(6)=0.240 eV$, and $\Delta \epsilon=0.005 eV$ for $7 \leq s
\leq 14$.
Using this model in their rate equation approach, Rosenfeld
et al. find a
strong tendency for Heptamer formation at $T \geq 400K$. On
the other hand,
our KMC simulations, using the same model for $\epsilon
(s)$, did not reveal a
particular preference for {\em Heptamer} formation: mostly
larger compact
clusters were produced, with the number of constituent
atoms varying from
cluster to cluster. The difference must be attributed to
the mean field type
approximations inherent to the rate equation formalism.
 Having no reliable
kinetic means at our disposal to produce compact heptameric
islands, we used
only the simpler first model to provide realistic adsorbate
structures for our
scattering calculations.  The KMC simulations (for both
models) were performed
using the time-dependent-Monte-Carlo (TDMC) scheme,
suggested in \cite{Kang2}.
According to this scheme, instead of randomly chosing
particles and accepting
or rejecting moves according to the given transition
probabilities (as in
``traditional'' MC simulations), one performs a move in any
attempt, and
propagates the time accordingly. More explicitly, we first
calculate the
average transition rate $\langle r \rangle = 1/\langle
\omega_{i \rightarrow
f} \rangle$ out of state $i$, then randomly sample a given
$i \rightarrow f$
move with probability $\omega_{i \rightarrow f}/\sum
\omega_{i \rightarrow f}$
and, finally, perform this move and record the time elapsed
as $\Delta t =
1/\langle r \rangle$. Further details about this procedure
can be found
elsewhere \cite{Kang1,Kang2}.

\subsection{Interaction Potential of He with DLA
Structures}
\label{systems-HeDLA}

We proceeded to perform scattering calculations from the
DLA structures. Our
purpose here was to check the degree of specificity of the
features exhibited
by the angular intensity distribution obtained from the
Heptamer systems of
sections~\ref{systems-random1} and~\ref{systems-random2}.
The results of the
KMC grown structures are the best model system we had at
hand to simulate a
"non-Heptameric" epitaxial surface.  This model served as
our control-case,
against which the features identified for the Heptameric
models were tested.
We will consider more fully the question of what structural
features can be
extracted from the information contained in the scattering
results from the
DLA structures in a future article. The potential used for
the He scattering
calculations in this case was:

\begin{equation}
{V}({\bf r}) = {V_s}(z) + \sum_{i =
1}^{N}{V_{He-Pt}}(\left|{\bf r} - {\bf
r_i}\right|)
\end{equation}

where ${V_s}(z)$ and ${V_{He-Pt}}$ are given by
eqs. (\ref{eq:Vs},\ref{eq:VHePt}) respectively, and ${\bf
r_i}$ now runs
over the positions of all adsorbed Pt atoms.

\section{The Methods for the Scattering Calculations}
\label{methods}

We employed the sudden approximation
\cite{Benny:Sud1,Benny:Sud2,Benny:Sud3,Benny:review1,Benny:review2},
which has
proved very useful in studies of atom scattering from
defects.  Basically, the
sudden approximation requires that the momentum transfer in
parallel to the
surface be small compared with the momentum transfer normal
to the surface,
i.e \cite{Benny:Sud1,Benny:review2},

\begin{equation}
{\left|{\bf K'}-{\bf K}\right|} \ll 2k_z \label{eq:Sudcond}
\end{equation}

where $k_z$ is the incident wave number in the z direction,
${\bf K}$ is the
incident wave vector in parallel to the surface plane (zero
in our
simulations), and ${\bf K'}$ is any intermediate or final
wave vector in
parallel to the surface plane which plays a significant
role in the scattering
process. For a surface area A of atom type X which contains
a collection of
defects of the same atom type, the angular intensity
distribution is given by
the function

\begin{equation}
{P_{{\bf K}\rightarrow{\bf K'}}} = {1 \over {A^2}} \, {{
\left|\left( \int
d\mbox{\boldmath $\rho$} \,
            {e^{2\,i\,\eta (\mbox{\boldmath $\rho$}) +
i\,\mbox{\boldmath $\rho$} \cdot \left( {\bf K'} - {\bf K}
\right) }}
          \right)\right|}^2} \label{eq:PKK'}
\end{equation}

where $\mbox{\boldmath $\rho$}=(x,y)$ and
$\eta(\mbox{\boldmath
$\rho$})$ is the scattering phase shift computed for fixed
$\mbox{\boldmath $\rho$}$, given in the WKB approximation
by:

\begin{equation}
\eta(\mbox{\boldmath $\rho$}) = \int
_{{z_t}(\mbox{\boldmath $\rho$}
)}^{\infty}
     dz\,\left({\sqrt{{{{k_z}}^2} -
{{2\,m\,V(\mbox{\boldmath $\rho$} ,z)}\over
{{{{\hbar}}^2}}}}}-{k_z}\right) -
{k_z}\,{z_t}(\mbox{\boldmath $\rho$} )
\label{eq:eta}
\end{equation}

where $z_t(\mbox{\boldmath $\rho$})$ is the classical
turning point pertaining
to the integrand in eq.(\ref{eq:eta}). Periodic boundary
conditions were
employed. Condition (\ref{eq:Sudcond}) for the validity of
the sudden
approximation is expected to break down for systems of high
corrugation, and
an isolated adsorbate on an otherwise flat surface
generally represents, for
realistic parameters, a very substantial local corrugation.
Nevertheless,
previous calculations have shown that the sudden
approximation reproduces
rather well many features of the scattering from isolated
adsorbates
\cite{Benny:Sud2}.  Features for which it breaks down are,
e.g, intensity
peaks due to double collision events (in which the incoming
atom first hits
the surface and then the adsorbate or vice versa), which
are a particularly
sensitive manifestation of a strong corrugation (we note
that a
double-collision version of the Sudden approximation has
recently been
developed \cite{me:DC}).  The sudden approximation is thus
useful for the
interpretation of these features for which it works, and
offers useful
insights also at the points where it breaks down. The
advantage offered by the
simplicity of the sudden expression eq.(\ref{eq:PKK'}) is
particularly manifest
when considering a potential of the form:

\begin{equation}
V({\bf r}) = V(z-\xi({\bf R}))
\label{eq:Vxi}
\end{equation}

This form implies that the classical turning surface is a
translation of the shape function $\xi({\bf R})$
\cite{Benny:disorder}:

\begin{equation}
z_t({\bf R}) = \xi({\bf R})+z_0 \label{eq:zt}
\end{equation}

Here $z_0$ measures the distance of the repulsive
He-surface interaction from
some reference plane. The choice of $z_0$ does not affect
the physical
observables (i.e., the intensities). $z_0$ is determined by
the incidence
energy and angle. Thus eq.(\ref{eq:eta}) can be rewritten
as:

\begin{equation}
\eta({\bf R}) = \alpha-k_z(\xi({\bf R})+z_0)
\end{equation}

where $\alpha$ is a constant independent of ${\bf R}$.
 Hence
from eq.(\ref{eq:PKK'}):

\begin{equation}
P_{{\bf K}\rightarrow{\bf K'}} = {1 \over {A^2}} \,
{{\left|\left(\int d{\bf
R}\,{e^{i\,\left( -2\,{k_z}\,\xi({\bf R})  +
                  \left( {\bf K}'-{\bf K} \right) \cdot
{\bf R} \right)
}}\right)\right|}^2} \label{eq:P-with-xi}
\end{equation}

thus avoiding the need even for the numerical evaluation of
the
phase-shift eq.(\ref{eq:eta}). It should be noted that
within the sudden
approximation the potential eq.(\ref{eq:Vxi}) and a
hard-wall potential with
the
same classical turning surface are essentially equivalent.
Under certain
assumptions concerning the interaction potential, the
angular intensity
distribution contains all the information about the
scattering process.
Indeed, it has been shown by Gerber et
al. \cite{Benny:inversion1,Benny:inversion2} that for an
interaction potential of the form $V({\bf
r})=V_0(z)+V_1(z)\,Q(x,y)$, if
$V_0(z)$ is known, it is possible to determine the form of
$V_1(z)$ and
$Q(x,y)$ from the angular intensity distribution. This was
implemented with
good accuracy for a problem where several diffraction peaks
only are
important, for a crystalline surface.  However, in general
it is a complicated
experimental matter to measure the angular intensity
distribution over the
entire physical range, and one is usually restricted to
several
one-dimensional measurements.  Interpretation of the
angular intensity
distribution can very effectively be carried out by
analyzing intensity
maxima, which generally fall into three categories: (1)
Rainbow peaks
\cite{Benny:RBs}, related to scattering off inflexion
points of the classical
turning surface, (2) Fraunhofer peaks \cite{Lahee:Fr},
caused by interference
of waves traveling around a spherical object, and (3)
constructive
interference (Interference) peaks, a consequence of
interference of waves
scattered from a periodic structure of some sort (which
would be classified as
Bragg interference for a fully periodic surface).  The
position of each peak,
once identified, contains important structural information
about the surface.
Our ensuing analysis of the angular intensity distribution
will proceed along
these lines.  Another, particularly simple feature from the
experimental point
of view, is the attenuation of the specular peak $I$ by a
defect. This
attenuation is closely related to the cross-section of the
defect. The
cross-section may be extracted from eq.(\ref{eq:PKK'}) via
its experimental
definition \cite{Comsa:book}:

\begin{equation}
I = {{I_0}\, {{e^{-n\,\Sigma\,\theta }}}} \label{eq:comsa}
\end{equation}

where $I=P_{0 \rightarrow 0}$, $I_0$ represents the
specular scattering
intensity from the corrugated and flat (smooth,
defect-free) surfaces
respectively, $n$ is the number of defects per unit surface
area, and $\theta$
their coverage. The parameter $\Sigma$ in
eq.(\ref{eq:comsa}) can be
interpreted as the cross-section for scattering by a single
defect. When
viewed as a function of the incidence energy, $I$
oscillates with a frequency
that is characteristic of the height of the defect above
the surface. A
Fourier analysis of these oscillations reveals the
distribution of defect
heights, and if performed at a wide enough energy range,
will also reveal the
z-direction "fine-structure" of these defects.  This too
will be an approach
we shall implement in the next section. A combination of
angular intensity
distribution and $I$ attenuation analysis proves to be a
powerful tool in
extracting information about defect structure. This will be
developed in the
course of our analysis of the results, later in this
article.

\section{Results and Discussion}
\label{results}

The results will be outlined by considering separately each
of the models
described in section~\ref{systems}. Within each model, we
will point out the
main physical features present in the scattering
calculations, pertinent to
the classification of the type of disorder present on the
surface.  All
scattering calculations were performed within the sudden
approximation, with
the He impinging on Pt(111) at normal incidence.  As will
be demonstrated in
detail in the next sections, there are very dramatic
effects in the angular
intensity distribution.  For the case of the single
adsorbed Heptamer, nearly
all characterizing features of the Heptamer's geometrical
structure can be
discerned in an analysis of the intensity distribution. We
are able to
determine the symmetry, the linear extents, and the slope
and number of
inflexion points of the Heptameric shape function. Taken
together, these
define the Heptamer almost uniquely. For the case of a
distribution of
Heptamers adsorbed upon the surface, we see significant
differences in
the scattering results, between the model of
translationally random Heptamers,
and the model of rotationally random Heptamers. Both
models, in turn, differ
most considerably from the case of DLA structures.
Furthermore, we are able to
quantify the differences among the various cases, relying
upon the analysis of
the single Heptamer. Thus we offer {\em signatures} of
disorder structure
based on the analysis of a He scattering experiment. We
believe that the great
majority of the pertinent features we observe will be
experimentally
observable. Much like in the case of scattering from
ordered surfaces, an
analysis of the angular intensity distribution, albeit
considerably more
complicated, may serve to almost fully classify the type of
disorder present
on a disordered surface. We proceed to demonstrate how such
an analysis may
be carried out.

\subsection{Physical Features in the Single Heptamer
Intensity Distribution}
\label{features}

Figure~\ref{fig:contour} shows the results of calculations
using the potential
discussed in section~\ref{systems-single}. The collision
wave-number of the
impinging He was $k_z = 3.0\, bohr^{-1}$.  The contour
surface shown in
figure~\ref{fig:contour} is the full 2D angular intensity
distribution, i.e
${\rm P_{{\bf K}\rightarrow{\bf K'}}}$ with the z-axis on a
logarithmic scale.
What is most striking about figure~\ref{fig:contour} is the
unambiguous
presence of structure in the intensity distribution: it has
a sharp and
well-defined angular appearance. The structure reflects the
six-fold symmetry
of the Heptamer, as indeed expected from a consideration of
eq.(\ref{eq:PKK'}). This property is easily shown to hold
for other types
of symmetry, i.e, the momentum-space view of the cluster
via the angular
intensity distribution retains the symmetry of the
real-space cluster
structure. This is a feature which should show up
unmistakably in the
appropriate experiment, if several on-plane directions are
probed. Thus a
small cluster on a surface has an experimental signature.

\subsubsection{Rainbow, Interference and Fraunhofer
Effects}
\label{Rainbows-etc}

In this section we will present in some detail an analysis
of the major
physical features present in the angular intensity
distribution. This
distribution is however much too detailed to give a full
analysis. Such an
analysis is definitely desirable to fully characterize the
type of disorder
present on the surface: in practice the angular intensity
distribution
contains enough information to leave no ambiguity. It will
however be clear
from the following, that even a partial analysis may lead
one a long way
toward an understanding of the disorder one is dealing
with. At the end of the
section we summarize the main findings.\\
Figure~\ref{fig:AID3} shows a section along the $\Delta
K_x$ axis (at $\Delta
K_y=0$) for scattering at a wave-number of $k_z = 3.0\,
bohr^{-1}$, with a
resolution of $0.05\, bohr^{-1}$ along the $\Delta K_x$
axis.  The angular
intensity
distribution in figure~\ref{fig:AID3} exhibits, in addition
to a large
specular peak, also a multitude of other peaks and "hidden"
peaks, the pattern
being obviously symmetric in the momentum transfer $\Delta
K_x = K_x'-K_x$
($K_x = 0$) direction.  In order to interpret the various
peaks we proceed
along the lines mentioned in the previous section, i.e we
attempt to classify
the peaks whenever possible as either Rainbows, Fraunhofers
or Interferences.

\paragraph{Rainbows:}
The take-home message from the ensuing analysis, is that
Rainbow peaks appear
as a dominant and experimentally observable feature in the
angular intensity
distribution of He scattered from a single compact cluster,
and that their
position contains critical information about the surface
morphology in the
vicinity of the cluster, and about the interaction
potential with the He.
Measurement of the Rainbow peaks alone, is sufficient to
calibrate important
points in a simple model potential, and is hence a powerful
tool in analysis
of the surface disorder. Following will be a brief reminder
of the origin and
physical significance of Rainbows (this subject has been
treated in much
detail elsewhere; see, e.g., \cite{Benny:review2}), and a
detailed analysis of
Rainbows-related features in our scattering results.\\ It
is useful to employ
a stationary phase, approximate evaluation of the sudden
scattering amplitude,
as in \cite{Benny:Sud2}. The crude quasi-classical
stationary phase limit of
eq.(\ref{eq:PKK'}) predicts a Rainbow whenever the
phase-shift has an
inflexion point \cite{Benny:Sud2}:

\begin{equation}
\eta ''(x) = 0 \label{eq:eta''}
\end{equation}

The $x_0$ satisfying this condition dominates the
scattering. In the classical
limit this should show up as a singularity in the angular
intensity
distribution at momentum transfer $\Delta K_x$ such that

\begin{equation}
\Delta K_x =  -2\,\eta '(x_0)
\label{eq:deltaKx}
\end{equation}

where $x_0$ is the stationary phase point of the integrand
eq.(\ref{eq:PKK'})
for
momentum transfer $\Delta K_x$. The singularity of the
crude classical
evaluation is smoothed into a finite peak in the more
refined sudden
calculation. Figure~\ref{fig:PS3} shows the phase-shift
function $\eta(x)$ at
$y=0$, along with its inflexion points for $k_z = 3.0\,
bohr^{-1}$. Due to the
inversion symmetry of the Heptamer about the y-z plane, the
inflexion points
appear as symmetric pairs along the x-axis. Each such pair
gives rise to
corresponding symmetrically located Rainbow peaks, as is
clear from
eq.(\ref{eq:deltaKx}). The
similarity between $\eta(x)$ and the Heptamer's
isopotential (fig.
\ref{fig:turning}) is understood
from eq.(\ref{eq:zt}). Based on this similarity, we can
identify the physical
origin of the Rainbow peaks: while it may be difficult to
demonstrate this for
a general potential (but immediate for, e.g.,
eq.(\ref{eq:Vxi})), it is clear
that in the present case the inflexion points
of the phase shift correspond uniquely to those of the
turning surface. From
figure~\ref{fig:PS3} it can be seen that at most
four distinct Rainbows peaks are expected, one for each
inflexion point. The
points numbered 1 and 2 are due to the Heptamer's top
corrugation, number 3
due to its side, and number 4 due to the corrugation
induced by the long-range
attractive part of the He-Pt interaction at the vicinity of
the surface. We
expect such inflexion points (and the resulting Rainbows)
to be widely present
for adsorbates on flat surfaces, regardless of the
quantitative details of the
potentials. In our case, the corrugation associated with
point 4 is too small
to enable R4 to be discerned from the specular peak. R3 is
clearly
associated with a double collision in which the He first
strikes the
Heptamer's side and then undergoes a second collision with
the flat Pt
surface. As mentioned before, such Double-Collision
Rainbows cannot be
described within the sudden approximation, because it does
not include
double-collision effects. However, using the quantitative
predictions of
eqs.(\ref{eq:eta''},\ref{eq:deltaKx}), {\bf {we identify
the peaks at $\Delta
Kx=\pm 0.33\,
bohr^{-1}$ and the hidden peaks at $\Delta Kx=\pm 0.13\,
bohr^{-1}$ as R1 and
R2 respectively}}. The peaks have an intensity of $10^{-4}$
and $5\cdot
10^{-3}$ respectively with respect to the specular, and
should be well within
experimental reach (the intensity predictions of the sudden
approximation for
RB peaks are generally in good agreement with the exact
quantum-mechanical
calculations; see e.g. \cite{Benny:Sud2}). We did not find
a point $x_0$ that
satisfies the Rainbow condition eq.(\ref{eq:eta''}), i.e,
inflexion points of
the phase shift, for the maxima at other $\Delta K_x$
values. We found it
instructive to compare our results with those obtained by
scattering from the
same system, at a significantly lower incidence energy. Our
motivation was the
expectation that at a lower energy, the He probes the Pt
system at a more
global scale, and is hence more sensitive to the overall
structure of the
defect and less to local details. This follows also from
the reduced
penetration depth, resulting in a reduced effective
corrugation. The
scattering calculations were repeated for the He impinging
on the Pt surface
at normal incidence, with a wavenumber of $k_z = 2.0\,
bohr^{-1}$.
Figure~\ref{fig:AID2} shows the same section as
figure~\ref{fig:AID3} for this
energy, with a more than 4-fold improved resolution of
$0.012\, bohr^{-1}$
(compared to the scattering at $k_z = 3.0\, bohr^{-1}$).
Figure~\ref{fig:PS2}
shows the corresponding phase-shift at $y=0$, along with
its inflexion points.
The inflexion points at the Heptamer's top do not persist,
as at this low
incidence energy, the He particle sees a higher Coulomb
barrier in between the
top Pt nuclei. Hence it is less effective in penetrating
this "soft" region,
resulting in the smoothed appearance of the Heptamer's top.
The absence of
inflexion points does not, however, rule out the existence
of Rainbows: it
must be remembered that conditions (\ref{eq:eta''}) and
(\ref{eq:deltaKx})
follow from a crude stationary-phase analysis and are to be
considered merely
as sufficient conditions for Rainbows. We do however expect
a significant
reduction in the intensity of the Rainbows. Furthermore,
analysis of scattering
results at $k_z=4.0$ and $5.0\, bohr^{-1}$ (not presented
here) reveals that
the position of the Rainbows is nearly constant, showing
only a slight
distancing
from the specular peak with increasing kz, as expected from
the corresponding
increase in corrugation.  Returning to
figure~\ref{fig:AID2} we thus identify
the peak at $\Delta K_x=0.15\, bohr^{-1}$ as R2. Compared
with
figure~\ref{fig:AID3}, its intensity is seen to have indeed
decreased by as
much as two orders of magnitude. Due to the improved
$\Delta K_x$ resolution
we are
now also able to discern the hidden peak at $\Delta
K_x=0.018\, bohr^{-1}$ as
R4,
resulting from the long-range attractive part near the edge
of the Heptamer.
R1 is missing, apparently masked by other peaks.

\paragraph{Interference Peaks:}
The origin of these peaks is the constructive interference
between He waves
scattered from different atoms. On a perfect crystalline
surface, these would
be referred to as Bragg interference peaks. Here, we expect
them to arise from
constructive contributions from neighboring Pt adatoms in
the Heptameric
cluster. No contribution can be expected from the surface,
as it was treated
as completely flat. This is a realistic assumption for Pt
(111), which has an
extremely smooth surface, whereas the attraction of the
Heptamer's top
electrons to the underlying surface will result in a
significant corrugation.
Interference peaks are expected at $\Delta K_x =
{{2\,n\,\pi }\over a}$,
$a=5.23\, bohr$ being the Pt unit cell distance, $n$ an
integer, for
scattering along the x-direction. However, it turns out
that even though these
peaks exist, they are of negligible magnitude ($<10^{-6}$
with respect to the
specular) and not within experimental realm. We shall
return to this point
when we discuss the more realistic system of the next
section.

\paragraph{Fraunhofer Peaks:}
Finally we attempted to assign the remaining peaks using
the Fraunhofer
diffraction model for scattering from a hard hemisphere on
a flat, hard
surface, given by Lahee et al. \cite{Lahee:Fr}. The
Fraunhofer diffraction
intensities given by this model are:

\begin{equation}
P(\theta) \propto {{\left|{J_1}(d\,k\,\sin (\theta ))\,\csc
(\theta
)\right|}^2}\, \left( 1 + \cos (\theta ) \right)^2
\label{eq:Fraunhofer}
\end{equation}

where $\theta$ is the sum of the incidence and scattering
angles (simply the
scattering angle in our case), $d$ the hard-sphere radius,
$k$ the collision
wave-number, and $J_1$ denotes the Bessel function of
$1^{st}$ order. Given
that eq.(\ref{eq:Fraunhofer}) is derived under assumptions
valid for the
scattering of light waves on small obstacles \cite{Faubel},
we did
not expect it to accurately predict the intensities of the
Fraunhofer peaks (see also ref. \cite{Benny:RBs}). Hence
rather than trying to
fit eq.(\ref{eq:Fraunhofer}) or its parameters to our data,
we
used a simple prediction resulting from
eq.(\ref{eq:Fraunhofer}): that the
spacing between Fraunhofer peaks as a function of $K_x=k
\sin (\theta)$ is
very nearly constant. We expect this property of
the Fraunhofer model to be more robust than the
intensities, as the latter are
modified by modulations due to other sources (such as
Rainbows). While
constant spacing cannot be claimed to be unique to the
Fraunhofer model, we
are not aware of any other relevant model with the same
property, which is in
addition appealingly simple. Thus we tried to fit a linear
regression for the
spacings of the remaining peaks. This could be done with a
maximum error of
6\% per peak and a coefficient of determination $R^2=0.995$
(fig. \ref{fig:Fr-constant}). The constant
peak-spacing prediction is expected to persist along the
principal axes also,
with a more sophisticated two-parameter model of scattering
from a hard
ellipsoid replacing the hemisphere. Indeed, we find a value
of $\Delta
Ky(F)/\Delta Kx(F)=1.16$ for the ratio of the predicted
Fraunhofer peak
spacings in the $\Delta K_y$ and $\Delta K_x$ directions,
compared to a
geometric ratio of
$R_x/R_y=1/\cos (\pi/6)=1.15$ between the major and minor
axes of the Heptamer
(the inversion is expected due to the conjugate
relationship between real and
Fourier space). We thus confidently identify the remaining
peaks as Fraunhofer
diffraction peaks.  Once again, it was instructive to
compare these results to
the case of a lower incidence energy. Most of the structure
in~\ref{fig:AID2},
at $k_z=2.0 bohr^{-1}$, is in fact due to Fraunhofer peaks.
Indeed, the very
regular appearance of the non-rainbow peaks and their
almost smooth envelope
is very reminiscent of a Fraunhofer interference pattern.
Relying on the
constant spacing property of eq.(\ref{eq:Fraunhofer})
again, we were able
to fit a linear regression to the peak spacings with a
maximum error of 1.3\%
per peak and a coefficient of determination $R^2=0.99995$
(fig. \ref{fig:Fr-constant}). The dominance and high
regularity of the
Fraunhofer pattern at this energy is explained by the
degree of smoothness of
the Heptamer itself, in addition to it being a single
corrugation on an
otherwise flat surface.  As we shall see in the next
section, this feature
breaks down when many adsorbates are present on the
surface. The Fraunhofer
pattern also obscures any Interferences, which were in any
case expected to be
small due to the low corrugation of the Heptamer's top.

\subsubsection{The Cross-Section of a Single Heptamer}

The cross-section, being the simplest parameter to extract
experimentally
(indeed, the bulk of the experiments done so far in He
scattering have been
cross-section measurements. See, e.g, ref.
\cite{Comsa:book}), is probably
also the
single most descriptive parameter of the surface disorder.
As we shall see in
this section, the cross-section has an oscillatory behavior
in the incidence
energy, and there is a mathematical procedure, outlined in
appendix A,
which enables a determination of the cluster heights
distribution from these
oscillations alone. However, it should be realized that the
necessary large
scale variations of the incidence energy are not easily
realizable
experimentally.
\\ Figure~\ref{fig:cs} shows the cross-section for He
scattering at normal incidence of Pt Heptamers at energies
corresponding to
$k_z=1, \ldots ,4\, bohr^{-1}$. The cross-section was
calculated from
eq.(\ref{eq:comsa}), which reads:

\begin{equation}
\Sigma = -{{\log ({{I}\over {{I_0}}})}\over {n\,\theta }}
\end{equation}

when inverted for $\Sigma$. In our case $n\,
\theta=1/(total\, area)$ (there
is one defect - the Heptamer - on the entire surface) and
$I_0=1$ (the
defect-free surface is flat). Two main points are worth
noting from
figure~\ref{fig:cs}:
\begin{enumerate}
\item{
\flushleft{\em The value of the cross-section, which lies
between 320 and 430
$\AA^2$}:\\
The cross-section of single Pt adatoms on Pt(111) has not
been measured yet.
We assume it to be similar to that measured for a number of
adsorbed molecules
and for vacancies, i.e about 120 $\AA^2$ for a room
temperature He beam
incident at $\theta_{inc}=40^{\circ}$ \cite{Comsa:book}.
This then yields an
effective Pt radius of $9 \AA$, and, using the
superposition principle
\cite{Comsa:book} together with a Pt unit cell distance of
$2.77 \AA$, we
arrive at a cross-section of about $\pi(2.77+9)^2
\cos(40^\circ)=195 \AA^2$
for He
scattering off a Pt Heptamer at normal incidence. Thus our
calculated
cross-section is about twice as large as the empirical
estimate. This almost
certainly reflects the inaccuracy in our value for $c_6$,
as explained in
section~\ref{systems-single}. However, the actual
cross-section of a Pt atom
on Pt(111) still remains to be measured and it may come out
higher than the
value employed above. At any rate, our main purpose in this
article is to
provide qualitative insight into the scattering features.
We thus note that it
is important to obtain a reliable experimental value for
the single adatom
cross-section, which once known, can be used to estimate
the cluster
cross-section. This in turn, can narrow down strongly the
number of possible
geometric configurations to be considered in attempting to
identify the
cluster type.
}
\item{
\flushleft{\em The cross-section oscillations}:\\
These arise from interference between waves scattered from
the flat part of
the Pt surface and the Heptamer's top. Clearly the period
of oscillations is
related to the height difference between the surface and
the top of the
corrugation. A Fourier analysis of the oscillations in
figure~\ref{fig:cs}
is shown in figure~\ref{fig:fourier}, and yields a height
difference of about
$5.5\, bohr$. This compares quite well
with a direct calculation from the classical turning
surface (see
figure~\ref{fig:turning}), which yields an average height
of about $6.5\,
bohr$. In the Heptamer case there are effectively two
heights to be considered
(marked $h_1$, $h_2$ in figure~\ref{fig:turning}). To see
the "fine structure"
($h_2$) from the cross-section oscillations, clearly a
large range of
incidence energies must be employed. In the case of the
Heptamer, our
calculations show an unrealistic several hundreds of eV
would be required to
discern $h_2$.  However, with adsorbates composed of
several atom types, or
even an adsorbate of different substance than the surface
(like in the
adsorbate-induced restructuring of surfaces of Cu(110)-($2
\times 1$) reported
by Kern et al. \cite{Kern}), the situation could be quite
different and
several experimentally observable heights could exist
within one cluster. A
theory explaining how to obtain the distribution of cluster
heights from the
cross-section oscillations is hence desirable, and appendix
A provides
a detailed calculation of this type in the case of a
$V=V(z-\xi({\bf R}))$
potential.
}
\end{enumerate}

\subsubsection{Summary of Results for Single Heptamer}
We sum up now briefly the main findings of this section:
\begin{enumerate}
\item{
Rainbow, Fraunhofer and Interference peaks are all expected
to be seen with
varying intensities in scattering from a Heptamer in the
same collision
conditions
(i.e the same experiment).
}
\item{
\flushleft
The Rainbow peaks due to the Heptamer's top corrugation are
expected to be the
most pronounced feature and to lie closest to the specular.
Fraunhofer peaks,
probing the size of the adsorbate, will vary in intensity
and the first few
may be masked by the much stronger Rainbows, however they
are expected to be
found at nearly constant spacing. Interference peaks due to
constructive
interference between waves scattered from the Heptamer's
top atoms may be too
weak to be measured experimentally.
}
\item{
\flushleft
As the incident He's energy is lowered, we expect the
angular intensity
distribution to become more regular, with the Fraunhofer
peaks gradually
overtaking the Rainbow peaks in dominance. This reflects
the prevailing of the
small-scale corrugation at high energies, which is smoothed
out when the De
Broglie wavelength of the incident He becomes more
comparable with the
Heptamer's dimensions at $k_z \leq 1.2\, bohr^{-1}$.
}
\item{
\flushleft
The cross-section is a most useful tool in that it is both
very simple to
obtain experimentally and also provides very important
information about the
defect's size and topography. The size information can be
used to narrow down
the possible candidates for cluster shapes, and the
topography provides an
excellent characterization of the defect's appearance.
}
\item{
\flushleft
As each of the scattering features discussed here is
sensitive to different
aspects of the interaction potential, experiments measuring
all these features
should provide a nearly complete signature of a Heptamer,
if present on a
surface. This conclusion should hold quite generally for
other adsorbate
shapes. For example in the Heptamer case, the prevailing of
Fraunhofer peaks
at low incidence energy suggests one is dealing with a
nearly hemispherical,
i.e. compact object. One may then combine the measured
cross-section and the
diameter estimated from the Fraunhofer peaks' position to
calculate how many
Pt atoms participate in a given cluster. The number of
observable Rainbow
peaks gives one a lower bound on the number of distinct
inflexion points in
the cluster, which can be used to geometrically reconstruct
the cluster. The
positions of the Rainbow peaks tell one about the degree of
corrugation of the
cluster. An analysis of the cross-section oscillations
reveals the number of
different heights in the cluster, their values and relative
frequency of
appearance. Taken together, these data combine to determine
the cluster quite
fully.
}
\end{enumerate}
We proceed now to examine the more realistic model of
scattering from a
collection of randomly adsorbed Heptamers.

\subsection{Scattering Results from Randomly Adsorbed
Heptamers and DLA
Structures}
\label{results-random}

In the following sections we analyze in some detail the
results for scattering
from the randomly adsorbed Heptamers and the DLA
structures. The main message
here, is that structural features persist in the angular
intensity pattern of
distributions of compact clusters, and can be quantified to
extract
interesting parameters of the disorder. Very little
structure, if any, however,
persists in the intensity distribution of DLA, in sharp
contrast to the
compact cluster case. Thus a scattering experiment can
readily distinguish
between compact clusters and DLA adsorbed on a surface.\\
We proceed to analyze our results. A summary is presented
at the end of the
section.

\subsubsection{Translationally Random Heptamers on Pt(111)}
\label{results-random1}

Figure~\ref{fig:AIDoriented} shows the results of the
scattering calculations
from the system described in section~\ref{systems-random1},
again for
$k_z=2.0\, bohr^{-1}$.  There is a rich structure in this
angular intensity
distribution and we shall not attempt to assign every peak.
However, relying
on the analysis of the previous sections, we identify
Rainbow peaks at $\Delta
K_x=\pm 0.024\, bohr^{-1}$ (R4), $\Delta K_x=\pm 0.12\,
bohr^{-1}$ (R2), and
$\Delta K_x=\pm 0.36\, bohr^{-1}$ (R1). These values agree
very well with
those found in the single Heptamer case, and we attribute
the small
discrepancies chiefly to numerical deviations. The reversal
in relative
intensities of R4 and R2 (R2 appears more intense now) is
believed to be
the result of the different role played by the Fraunhofer
diffraction pattern
in the current system. The optimal conditions for
Fraunhofer diffraction are
when there is no interference between neighboring clusters,
as was certainly
the case for the single Heptamer. In the present system the
average distance
between Heptamer centers is rather small, giving rise to
significant
interference between neighboring Heptamers. As a
consequence the Fraunhofer
pattern is strongly attenuated. The increased intensity of
R4 in
figure~\ref{fig:AID2} is hence attributable to an
underlying Fraunhofer peak.
Considering the shallowness of the attractive well giving
rise to R4 (see
figures \ref{fig:PS3},{fig:PS2}), the "normal" state of
affairs is restored in
figure~\ref{fig:AIDoriented}, where R2 is the more intense.
The attenuation
of the Fraunhofer pattern allows for the reappearance of
R1, which was
missing (masked) in figure~\ref{fig:AID2}.
The strong Interference peaks at $\Delta Kx=\pm 1.2\,
bohr^{-1}$ can of course
not go
unnoticed.  Their intensification must be attributed to the
constructive
contributions of all Heptamers, combined with the
diminished masking by the
Fraunhofer peaks.

\subsubsection{Translationally and Rotationally Random
Heptamers on Pt(111)}
\label{results-random2}

Figure~\ref{fig:AIDrandom} shows the results of the
scattering calculations
from the system described in section~\ref{systems-random2},
at the same
incidence wavenumber of $k_z=2.0\, bohr^{-1}$ as in the
previous section. It
is most instructive to compare figure~\ref{fig:AIDrandom}
to
figure~\ref{fig:AIDoriented}. Clearly the envelope of the
intensity spectrum
is much reduced, as expected from the disappearance of most
of the
constructive contributions when the Heptamers assume random
orientations. Yet
the Rainbow peaks persist, albeit at a lower intensity, and
we again identify
R4 at $\Delta K_x=\pm 0.024\, bohr^{-1}$, R2 at $\Delta
K_x=\pm 0.12\,
bohr^{-1}$ and R1 at $\Delta K_x=\pm 0.36\, bohr^{-1}$.
Like the Rainbows the
Interferences at $\Delta Kx=\pm 1.2\, bohr^{-1}$ are also
much weaker,
corresponding
to the fact that there are now fewer constructive
contributions in the
original $\Delta K_x$ direction, and most of the intensity
is present in other
directions.

\subsubsection{Scattering Pattern for DLA Structures}
\label{results-DLA}

The scattering results from this system, described in
section~\ref{systems-DLA} (first model), are presented in
figure~\ref{fig:AIDislands}. The incident He's wavenumber
was once again
$k_z=2.0\, bohr^{-1}$. The most
prominent feature in figure~\ref{fig:AIDislands} as
compared to
figures~\ref{fig:AIDoriented} and~\ref{fig:AIDrandom} is
the almost complete
disappearance of structure in the angular intensity
distribution. The peaks
that can still be discerned are the Interferences at
$\Delta K_x=\pm 1.2\,
bohr^{-1}$ and a new set at $\Delta K_x=\pm 0.096
bohr^{-1}$. The existence of
the small Interference peaks is explained by the hexagonal
unit cell structure
of the islands. This is of course far from a complete
hexagonal surface at the
5\% coverage of figure~\ref{fig:AIDislands}, but is
sufficient to produce some
constructive interference between neighboring Pt atoms. The
peaks at $\Delta
K_x=\pm 0.096 bohr^{-1}$ are not at the position of any of
the Rainbow peaks
from the previous two sections, but can still be considered
as Rainbows, this
time an R1 or R2 (they are identical from symmetry
considerations except for
at the rim of the islands).  They arise from the weak
corrugation at the top
of the monolayer islands.  Another dominant feature of
figure~\ref{fig:AIDislands} is the broad specular peak.
This means that
intensity is distributed almost continuously into nearly
specular directions.
We shall not attempt an analysis of the width of the
specular peak and its
behavior as a function of $\Delta K_x$ here. We shall limit
ourselves to
mentioning that these features are related to the fractal
dimensions of the Pt
islands \cite{Pfeifer:fractals}. A detailed examination of
this issue in the
context of He scattering will be presented in a future
article (see also ref.
\cite{Ernst:3}).  Needless to say, no Fraunhofer peaks
pattern is observable
in figure~\ref{fig:AIDislands}.

\subsubsection{Summary of Results for Randomly Adsorbed
Heptamers and DLA
Structures}

To summarize our findings in the last sections:
\begin{enumerate}
\item{
There is a rich structure of several peaks in the case of
He scattering off
randomly adsorbed Heptamers. Much of this structure
persists, albeit at a
lower intensity, for the case of randomly rotated
Heptamers. However, in the
case of randomly structured islands, there is nearly no
trace of structure in
the angular intensity distribution.  Thus a powerful
preliminary criterion to
distinguish structured clusters from randomly shaped ones,
is the respective
structure of their angular intensity distribution.
}
\item{
The source of the structure in the angular intensity
distribution, in
the compact-cluster case, are primarily the Rainbow peaks.
These are at the
same time very robust to perturbations of the structure
(e.g a random rotation
of all clusters) and strongly position-related to the fine
details of the
cluster structure. Thus the Rainbow peaks may serve as a
crucial component in
the identification of surface structure.
}
\item{
At this relatively low concentration, the Fraunhofer peaks
are of secondary
importance. They are quite sensitive to adsorbate proximity
and shape.
}
\item{
Interference peaks tend to reflect the underlying surface
symmetry as
displayed by the adsorbed clusters and are hence not of
major help in
determining adsorbate structure.
}
\item{
A broad specular peak without much other structure is
indicative of randomly
shaped islands on the surface.
}
\item{
The question arises whether the identification of cluster
type is still as
clear as in the single Heptamer case. We believe the answer
to be affirmative.
The major difference between the single Heptamer and
randomly adsorbed
Heptamers cases was the prevailing of Fraunhofer peaks in
the former. This
resulted from the lack of interference between neighboring
clusters in the
single Heptamer case. This situation can however be
experimentally approached
by reducing adsorbate concentration. If compact clusters
are formed, they will
then clearly create a Fraunhofer diffraction pattern. With
increasing
concentration interference between neighboring clusters
will destroy the
Fraunhofer pattern, but if the average cross-section
remains the same, one may
conclude quite safely that compact clusters remain present.
If compact
clusters do not form at concentrations where there is
negligible interference
between neighboring clusters, one must resort to relying on
Rainbows and
cross-section analysis alone. However, as discussed in the
single Heptamer
case, this analysis can lead one a long way toward a
cluster signature.
}
\end{enumerate}

\section{Concluding Remarks}
\label{conclusions}

In this study we focused on the utility of TEAS as a tool
for learning about
surface disorder. More specifically, we
sought to demonstrate that the angular intensity
distribution of He atoms
scattered from adsorbed clusters contains information which
can
be used to determine most, if not all the characteristics
necessary for a
signature of a cluster. This determination is based on the
identification of
Rainbow and Fraunhofer peaks in the angular intensity
distribution, and an
analysis of cross-section oscillations dependence on
incidence energy.  All
these features, by probing different aspects of the
interaction of the
adsorbed clusters and the colliding atoms, when taken
together, combine to
provide an almost complete identification of the cluster.
Although our results
were obtained for the particular case of Heptameric
clusters, common physical
sense dictates that the qualitative features should persist
for other cluster
types. That is, as we have demonstrated for Heptameric
clusters, sharp and
well-defined characteristics should be extractable from the
angular intensity
distribution. This should certainly be so for the case of a
single cluster (or
a dilute concentration of clusters), and with some more
labor for the case of
a distribution of clusters. For example, we expect that in
the case of a
distribution of compact trimers, only one Rainbow peak will
be seen, the
Fraunhofer peak spacing will reflect the structure of the
underlying
triangular base, and the cross-section will indicate the
presence of three
atoms. Thus the scattering can serve as a
crystallographic tool for disordered surfaces. The major
reservation regarding
the last statement is, that a distribution of sizes and
shapes of clusters,
may obscure the otherwise sharp features present in the
diffraction
spectrum. Viewed from another
perspective, as we have shown for the case of
translationally random Heptamers
versus translationally and rotationally random Heptamers,
versus DLA,
inspection of the angular intensity distribution is a means
to
distinguish one type of disorder from another. We are aware
that the effects
discussed here are hardly easily detectable. Yet, we have
performed our
calculations in a realistic energy regime, and by our
estimates most of them
should be observable with present day techniques and
resolutions.  On this
basis one may anticipate considerable advance in knowledge
of compact clusters
adsorbed on surfaces and their interaction with incoming
atoms, once the
pertinent experiments are performed. An issue we have
however only scratched
the surface of, is that of scattering from DLA.  Whereas we
have shown that
the structure of the angular intensity distribution in this
case is
substantially different from the compact clusters case, and
hence such
structures are easily distinguished from compact clusters,
it is as of yet an
open question what information can be extracted from the
angular intensity
distribution in this case and how this information could be
used to
characterize such structures.  Work on this issue is in
progress and once
completed should significantly enhance our ability to
understand surface
clusters.
In light of the very impressive results obtained recently
by means of the STM
technique (see, e.g., \cite{Comsa:STM2,Lagally:STM}), a few
comments on the
relative position of the two techniques are in order. In
our view, TEAS and
STM are really {\em complementary} techniques, in the sense
that STM probes
the local structure (several millions of $\AA^2$), whereas
TEAS probes the
global structure. In addition, TEAS is better at probing
defects with high
mobility, and does not damage the surface. However, there
is another
significant difference, which is frequently overlooked.
This is the fact that
STM mostly measures the electronic density near the Fermi
level, whereas He
scattering is sensitive to the {\em total} electronic
density, via the
He-surface interaction potential \cite{Norskov}. Thus
although He scattering
suffers from the significant disadvantage of displaying the
surface in
momentum space, whereas STM shows the surface in real
space, He scattering is
applicable to a much wider range of systems (including
liquids
\cite{Benny:liquids}). STM, on the other hand, can be
applied only to metal
surfaces. We hope to have convincingly demonstrated that
even in a field in
which STM has taken a definite lead, He scattering will
have much to offer
once the pertinent experiments are performed.

\noindent {\bf Acknowledgements}\\
This research was supported by the German-Israeli
Foundation for Scientific
Research (G.I.F.), under grant number I-215-006.5/91 (to
R.B.G). Part of this
work was carried out with support from the Institute of
Surface and Interface
Science (ISIS) at UC Irvine. The Fritz Haber Center at the
Hebrew University
is supported by the Minerva Gesellschaft f\"{u}r die
Forschung, Munich,
Germany. We are very grateful to Profs. G. Comsa, B.
Poelsema and I.
Steinberger, and Dr. P. Zeppenfeld for many discussions
that stimulated much
of this work.

\newpage

\appendix{Appendix \protect\newline{}}
{\bf Calculation of Surface Heights Distribution from
Specular Peak
Attenuation:}\\

\noindent It suffices to measure the intensity of the
specular peak over a
range of incidence energies to obtain the distribution of
surface heights.
Consider eq.(\ref{eq:P-with-xi}), which was obtained under
the assumption of a
potential of the form (eq.(\ref{eq:Vxi})):\\

$V({\bf r}) = V(z-\xi({\bf R}))$  \\

Assuming a specular collision (${\bf K'}={\bf K}$), we
expand
eq.(\ref{eq:P-with-xi}) to obtain:

\begin{equation}
{I_0}(k_z) \equiv {P_{{\bf K} \rightarrow {\bf K}}} = {1
\over {A^2}} \, { \int
\int d\mbox{\boldmath $R$} \, d\mbox{\boldmath $R'$}
            {e^{2\,i\,k_z\,[\xi (\mbox{\boldmath $R'$}) \,
- \,\xi (\mbox{\boldmath $R$})]}}} \label{eq:I0}
\end{equation}

We know that the cross-section, and hence the specular peak
intensity,
oscillates with the incidence energy. To see this, we
Fourier transform
${I_0}(k_z)$. Defining first the difference between shape
functions:

\begin{equation}
\Theta({\bf R},{\bf R'})=-2 \, [ \, \xi({\bf R}) \,- \,
\xi({\bf R'}) \, ] \label{eq:theta}
\end{equation}

the Fourier transform is given by:
\begin{eqnarray}
{\cal I}(z)={\cal F}[{I_0}(k_z)]={1 \over {\sqrt{2 \pi} \,
A^2}} \, { \int \int
d\mbox{\boldmath $R$} \, d\mbox{\boldmath $R'$} \: \int
d{k_z} \, e^{i \, k_z
\, [\, z \, - \, \Theta({\bf R},{\bf R'}) \, ]}} \nonumber
\\
={\sqrt{2 \pi} \over {A^2}} \, {\int \int d\mbox{\boldmath
$R$} \,
d\mbox{\boldmath $R'$} \, \delta [\Theta({\bf R},{\bf R'})
\, - \, z]}
\label{eq:Iz}
\end{eqnarray}

Equation (\ref{eq:Iz}) can in principle be solved for
${\cal I}(z)$, given the
shape function ${\xi ({\bf R})}$. We claim that its
normalized value for a
given height above the surface $z$, is the relative
frequency of appearance of
that height. In other words, ${\cal I}(z)$ is the
distribution of surface
heights.  To see this, consider the following example,
corresponding to a
radically simplified ``lego''-model of 3D epitaxial growth.
Suppose that atoms
can adsorb only along the x-axis, and only one
nucleation-site is available.
Suppose further that the resulting defect's shape-function
is adequately
represented by a stacking of rectangles of width $w$ and
height $h$. The
situation is represented schematically in
figure~\ref{fig:lego}. Let the number
of atoms making up the base of the defect, i.e, the total
number of heights be
$N$. We denote the height at $x_j$ by $h_j$. Let the number
of {\em distinct} heights be $M$, i.e, there are $M$
different heights ${ \{ l_i \} }_{i=1}^M$, each appearing
$n_i$ times. Thus:

\begin{equation}
N = {L \over w} = \sum_{i=1}^M n_i \label{eq:N}
\end{equation}
Next define a height-function:

\begin{equation}
{{\Delta}_j}(x) = \left\{ \begin{array}{ll}
	0       & \mbox{if $x<j \, w$ or $x > (j+1)\, w$}
\\
	-2{h_j} & \mbox{if $j \, w \leq x \leq (j+1)\, w$}
	\end{array}
\right.
\end{equation}

Using this, we have for given $x'$:

\begin{equation}
{\Theta}(x-x') = \left\{ \begin{array}{ll}
	 {{\Delta}_j}(x)-{{\Delta}_{j'}}(x') & \mbox{if $0
\leq x,x' \leq L$}\\
	 {{\Delta}_j}(x)                     & \mbox{if $0
\leq x \leq L$ and
$(x' \leq 0$ or $x' \geq L)$}\\
	 {-{\Delta}_{j'}}(x')                & \mbox{if $0
\leq x' \leq L$ and
$(x \leq 0$ or $x \geq L)$}\\
	 0                                   & \mbox{if
$x,x'<0$ or $x,x'>L$}\\
	\end{array}
\right.
\end{equation}

Now we can evaluate~\ref{eq:Iz}:
\begin{equation}
{\int dx' \int dx\: \delta [\Theta(x,x') \, - \, z]} =
{\int _{-\infty} ^{\infty} dx'\: \left[ I_1\: +\: I_2\: +\:
I_3   \right]}
\end{equation}

where:

\begin{equation}
I_1={\int_0^L dx\: \delta \left[ \left\{
	\begin{array}{ll}
	 {{\Delta}_j}(x)-{{\Delta}_{j'}}(x')-z & \mbox{if
$0 \leq x' \leq L$}\\
	 {{\Delta}_j}(x)-z & \mbox{if $x'<0$ or $x'>L$}
	\end{array} \right\} \right] }
\end{equation}

\begin{equation}
I_2={ \int _{-\infty} ^0 dx\: \delta \left[ \left\{
	\begin{array}{ll}
	 {-{\Delta}_{j'}}(x')-z & \mbox{if $0 \leq x' \leq
L$} \\
	 -z                     & \mbox{if $x'<0$ or
$x'>L$} \\
	\end{array} \right\} \right] }
\end{equation}

\begin{equation}
I_3={\int _L ^{\infty} dx\: \delta \left[ \left\{
	\begin{array}{ll}
	 {-{\Delta}_{j'}}(x')-z  & \mbox{if $0 \leq x' \leq
L$} \\
	 -z                      & \mbox{if $x'<0$ or
$x'>L$}  \\
	\end{array} \right\} \right] }
\end{equation}

Expansion of each term yields:
\begin{eqnarray}
\int I_1=
\overbrace{ \int_0^L dx' \int_0^L dx \: \delta [
\,{{\Delta}_j}(x)-{{\Delta}_{j'}}(x')-z\, ] }^{\rm III} \:
+ \nonumber  \\
\overbrace{ \int_0^L dx' \int_{-\infty}^0 dx \: \delta [
\,-{{\Delta}_{j'}}(x')-z\, ] }^{\rm II} \: + \nonumber  \\
\overbrace{ \int_0^L dx' \int_L^{\infty} dx \: \delta [
\,-{{\Delta}_{j'}}(x')-z\, ] }^{\rm II}
\end{eqnarray}

\begin{eqnarray}
\int I_2=
\overbrace{ \int_{-\infty}^0 dx' \int_0^L dx \: \delta [
\,{{\Delta}_j}(x)-z\, ] }^{\rm II} \: + \:
\overbrace{ \int_{-\infty}^0 dx' \int_{-\infty}^0 dx \:
\delta [ \,-z\, ] }^{\rm I} \: + \nonumber  \\
\overbrace{ \int_{-\infty}^0 dx' \int_L^{\infty} dx \:
\delta [ \,-z\, ] }^{\rm I}
\end{eqnarray}

\begin{eqnarray}
\int I_3=
\overbrace{ \int_L^{\infty} dx' \int_0^L dx \: \delta [
\,{{\Delta}_j}(x)-z\, ] }^{\rm II} \: + \:
\overbrace{ \int_L^{\infty} dx' \int_{-\infty}^0 dx \:
\delta [ \,-z\, ] }^{\rm I} \: + \nonumber  \\
\overbrace{ \int_L^{\infty} dx' \int_L^{\infty} dx \:
\delta [ \,-z\, ] }^{\rm I}
\end{eqnarray}

It is simple to check that all terms overbraced with the
same numeral are
indeed equal. Their evaluation yields:

\begin{equation}
{\rm  I}={(M-{L \over 2})}^2 \, \delta (z)
\end{equation}

\begin{equation}
{\rm  II}= \int_L^{\infty} dx' \sum_{j=0}^{N-1} \int_{j \,
w}^{(j+1)\, w} dx \,
\delta (-2\, h_{j}-z) = (M-{L \over 2})\, w
\sum_{j=0}^{N-1} \delta (2\, h_{j}+z)
\end{equation}

\begin{eqnarray}
{\rm  III}= \sum_{j'=0}^{N-1} \sum_{j=0}^{N-1} \int_{j' \,
w}^{(j'+1)w} dx'
\int_{j \, w}^{(j+1)w} dx\, \delta (-2\, h_j+2\, h_{j'}-z)
= \nonumber \\
w^2 \, \sum_{j'=0}^{N-1} \sum_{j=0}^N \delta
[2(h_{j'}-h_j)-z]
\end{eqnarray}

The defect-free zone is of length:
\begin{equation}
L_0 \equiv 2\, (M-{L \over 2}) \label{eq:L0}
\end{equation}

Collecting all terms and inserting the definitions
(\ref{eq:N}) and
(\ref{eq:L0}), we obtain finally the normalized result:

\begin{eqnarray}
{1 \over L_0^2}\, {\cal I}(z) = \underbrace{\delta(z)}_{\rm
I'} +
\underbrace{{2 \over N}\, {({L \over L_0})}\, \sum_{i=1}^M
{n_i}\, \delta (z+2\, l_i)}_{\rm II'} + \nonumber \\
\underbrace{ {1 \over N^2}\, {({L \over L_0})}^2
\sum_{i'=1}^M \sum_{i=1}^M
{n_{i'}}\, {n_i}\, \delta [z-2(l_{i'}-l_i)] }_{\rm III'}
\label{eq:heights}
\end{eqnarray}

The physical significance of the terms in
eq.(~\ref{eq:heights}) is as follows:
\begin{enumerate}
\item{
${\rm I'}$ represents the contribution of the flat part of
the surface: a
unit-sized peak at $z=0$, the reference plane.
}
\item{
${\rm II'}$ is the most interesting term: it contains a
contribution from all
heights present on the surface $l_i$, each weighted by its
frequency of
appearance $n_i$. {\em The sum over these contributions is
the sought after
distribution of surface heights}. It is scaled as ${1 \over
N}\, {L \over
L_0}$.
}
\item{
${\rm III'}$ is a sum of peaks at all possible height {\em
differences}. This
information is redundant as it can be extracted from ${\rm
II'}$. It is scaled
as $({1 \over N})^2\, ({L \over L_0})^2$ and will therefore
be negligible compared
with ${\rm II'}$.
}
\end{enumerate}

Equation (\ref{eq:heights}) is a most useful result in our
opinion. It shows
that a very significant statistical feature of the
disorder, the distribution
of surface heights, can be extracted with ease from a He
scattering
experiment. Experimentally, all that needs to be done is to
measure the
specular scattering at a wide incidence energy range. A
Fourier transform will
reveal the desired distribution. \\
The situation for the more realistic situation of defects
on a surface (as opposed to a line) is even better:
our calculations show that the ``noise'' term ${\rm III'}$
is scaled down by a
factor of $({1 \over N})^2\, ({L \over L_0})^2$ with
respect to the ``signal''
term ${\rm II'}$.

\newpage

\noindent
\begin{figure}
\caption{
Schematic representation of translationally
random Heptamers at 5\% coverage on Pt(111). Although the
Heptamers occupy
random lattice sites, they clearly have orientations
compatible with the
underlying hexagonal symmetry of the surface.}
\label{fig:oriented}
\end{figure}

\noindent
\begin{figure}
\caption{
Translationally and rotationally random Heptamers
at 5\% coverage on Pt(111). Here the Heptamers do not obey
an orientational
constraint. Instead their orientation is picked at random
from a uniform
distribution. Their centers of mass occupy random lattice
sites.}
\label{fig:random}
\end{figure}

\noindent
\begin{figure}
\caption{
Pt islands produced by KMC simulations, at 5\%
coverage on surfaces of 100*100 unit cells, according to
the model with energy
independent of cluster size (first model; see text). The
surface is a
parallelogram with an angle of $60^{\circ}$. Periodic
boundary conditions are
employed, so straight edges usually imply that the cluster
is continued at the
opposite boundary. The configurations displayed differ only
in the random
number sequences used to generate them.}
\label{fig:islands}
\end{figure}

\noindent
\begin{figure}
\caption{
The full angular intensity distribution for He
scattering from a Pt Heptamer on a Pt surface, at
\mbox{$k_z = 3.0
bohr^{-1}$}. Axes: \mbox{$\Delta K_x$} (horizontal),
\mbox{$\Delta K_y$}
(vertical), log intensity (out of page). The six-fold
symmetry of the Heptamer
is clearly maintained in momentum space.}
\label{fig:contour}
\end{figure}

\noindent
\begin{figure}
\caption{
Log He scattering intensity as a function of
momentum transfer in x direction, at \mbox{$k_z = 3.0
bohr^{-1}$}. Positions
of rainbow peaks are indicated.}
\label{fig:AID3}
\end{figure}

\noindent
\begin{figure}
\caption{
Phase-shift function \mbox{$ \eta (x) $} as defined
in eq.(\ref{eq:eta}), at \mbox{$k_z = 3.0 bohr^{-1}$}:
section along x axis.
The inflexion points are indicated.}
\label{fig:PS3}
\end{figure}

\noindent
\begin{figure}
\caption{
Log He scattering intensity as a function of
momentum transfer in x direction, at \mbox{$k_z = 2.0
bohr^{-1}$}. Positions
of rainbow peaks are indicated. A strong Fraunhofer
interference pattern is
visible.}
\label{fig:AID2}
\end{figure}

\noindent
\begin{figure}
\caption{
Constant spacing between peaks in Fraunhofer
model: the positions of peaks identified as Fraunhofers in
the scattering
calculations are displayed as a function of peak number,
for two incidence
energies. The (nearly) constant spacing between peaks is
predicted by
eq.(\protect\ref{eq:Fraunhofer}).}
\label{fig:Fr-constant}
\end{figure}

\noindent
\begin{figure}
\caption{
Phase-shift function \mbox{$ \eta (x) $} at
\mbox{$k_z = 2.0 bohr^{-1}$}: section along x axis. The
inflexion points are
indicated.}
\label{fig:PS2}
\end{figure}

\noindent
\begin{figure}
\caption{
Total cross-section (in $\AA^2$) of the adsorbed
Heptamer, as a function of He incidence wavenumber
\mbox{$k_z (bohr^{-1})$}.}
\label{fig:cs}
\end{figure}

\noindent
\begin{figure}
\caption{
Fourier transform of the specular intensity versus
the height above the surface. The peak at $z \approx 10.5
bohr$ corresponds to
a distance of $\approx 5.5 bohr$ from the Heptamer's top to
the surface, as
explained in appendix A.}
\label{fig:fourier}
\end{figure}

\noindent
\begin{figure}
\caption{
Heptamer classical turning surface at y=0 for two
He incidence wavenumbers: \mbox{$k_z = 4.0 bohr^{-1}$} and
\mbox{$k_z = 1.7
bohr^{-1}$}. The two heights relevant to the distribution
of surface heights
are indicated.}
\label{fig:turning}
\end{figure}

\noindent
\begin{figure}
\caption{
He scattering intensity distribution from
translationally random Heptamers, along {$\Delta K_x
[bohr^{-1}]$} axis. He
incident at \mbox{$k_z = 2.0 bohr^{-1}$}. Rainbow (R) and
constructive
interference (Int.) peaks are indicated.}
\label{fig:AIDoriented}
\end{figure}

\noindent
\begin{figure}
\caption{
He scattering intensity distribution from
translationally {\em and} rotationally random Heptamers,
along {$\Delta K_x
[bohr^{-1}]$} axis. He incident at \mbox{$k_z = 2.0
bohr^{-1}$}.}
\label{fig:AIDrandom}
\end{figure}

\noindent
\begin{figure}
\caption{
He scattering intensity distribution from DLA,
along {$\Delta K_x [bohr^{-1}]$} axis. He incident at
\mbox{$k_z = 2.0
bohr^{-1}$}.}
\label{fig:AIDislands}
\end{figure}

\noindent
\begin{figure}
\caption{
``Lego'' model of 3D epitaxial growth. Schematically
drawn is the shape function of the surface with a single
defect adsorbed upon
it.}
\label{fig:lego}
\end{figure}

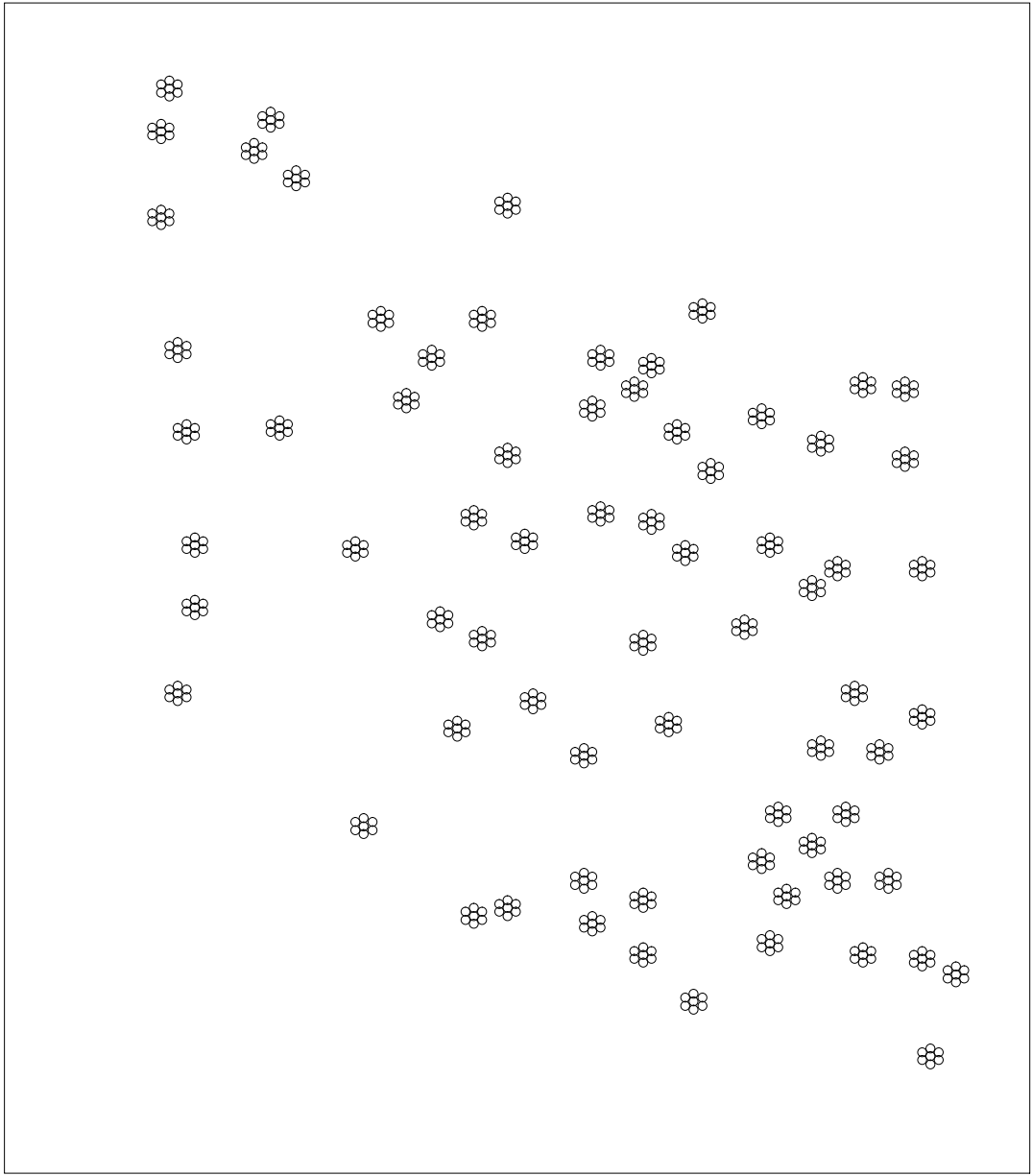
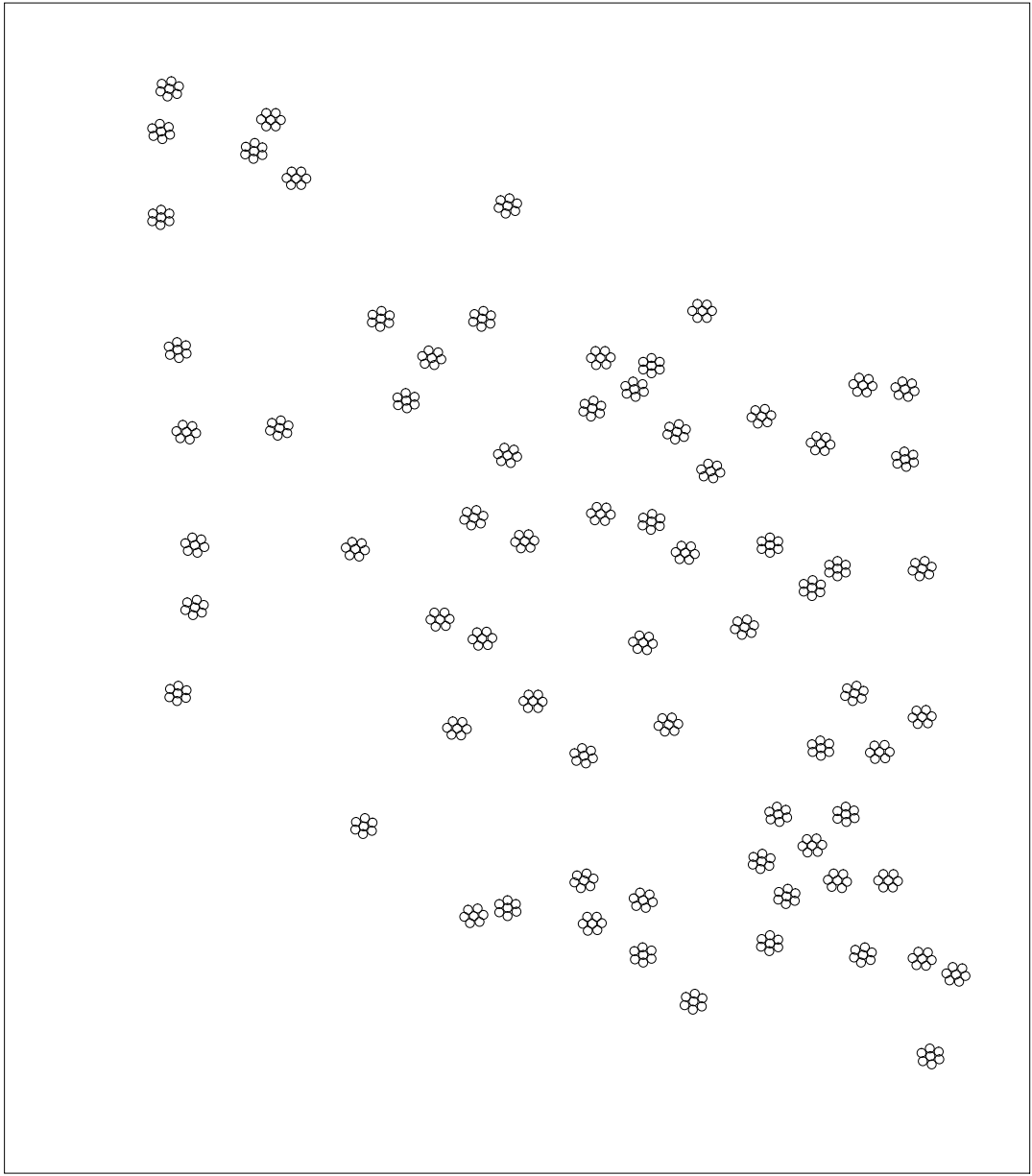
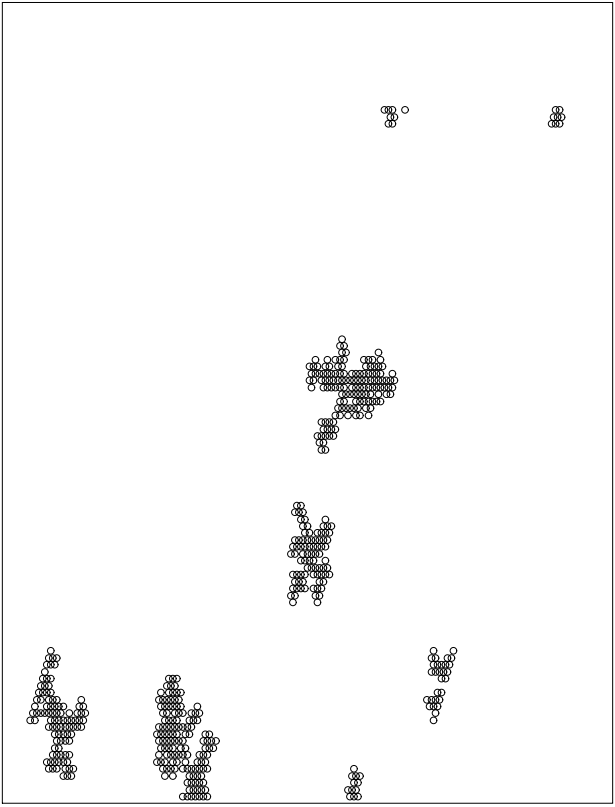
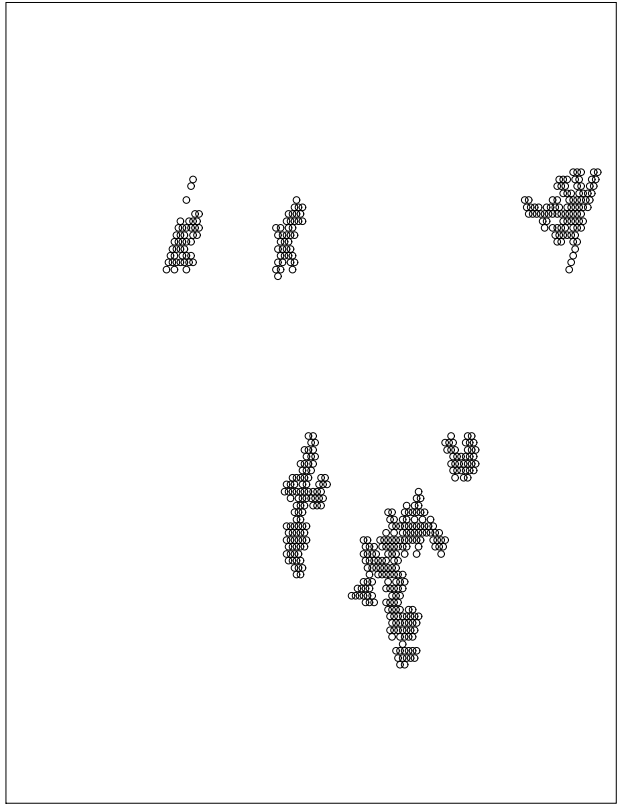
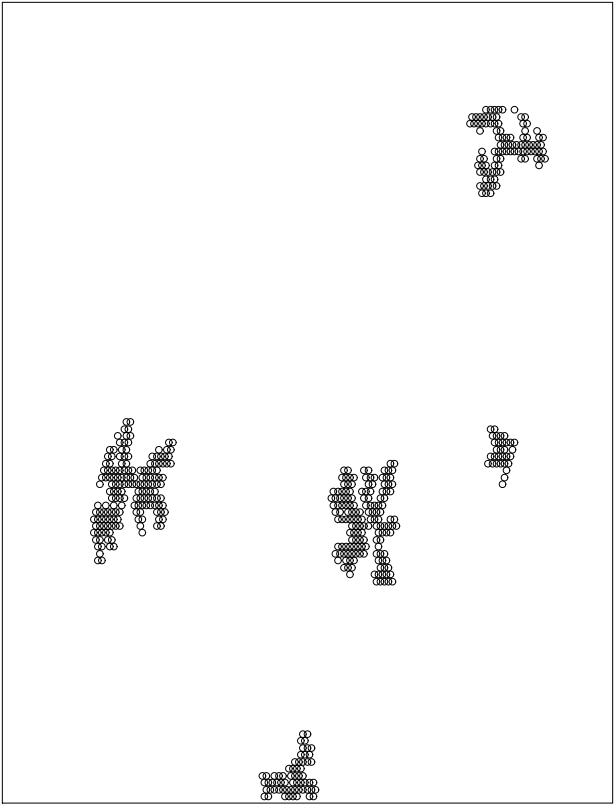
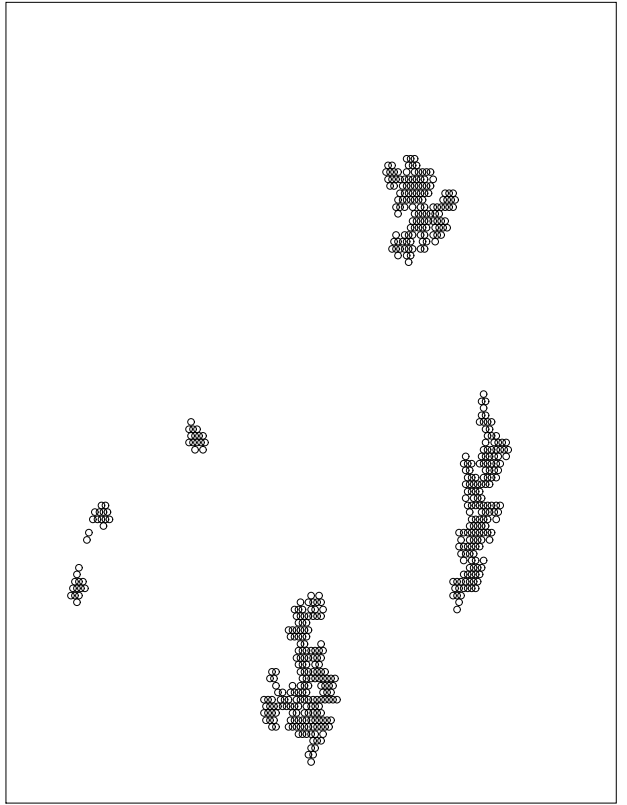
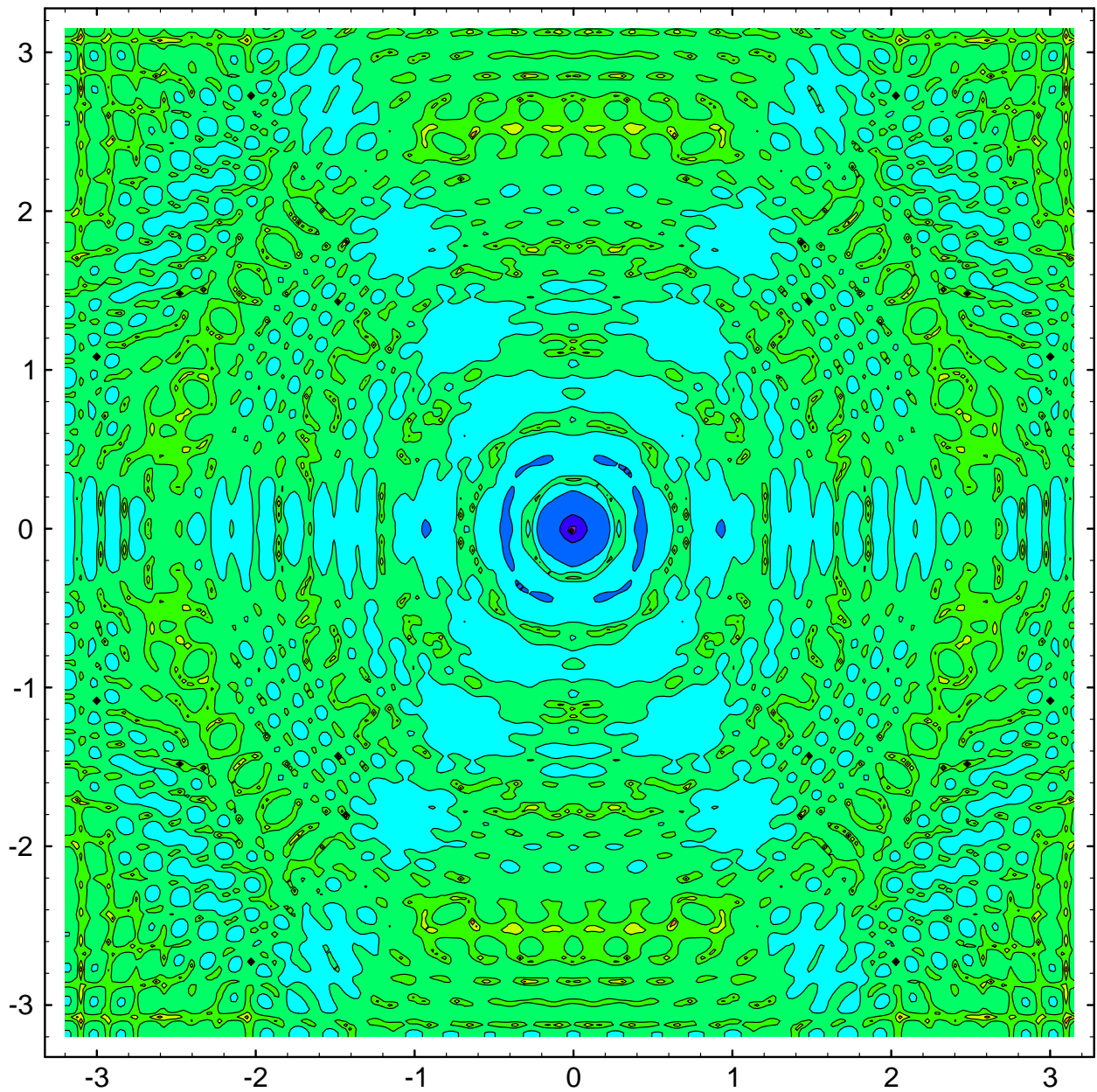
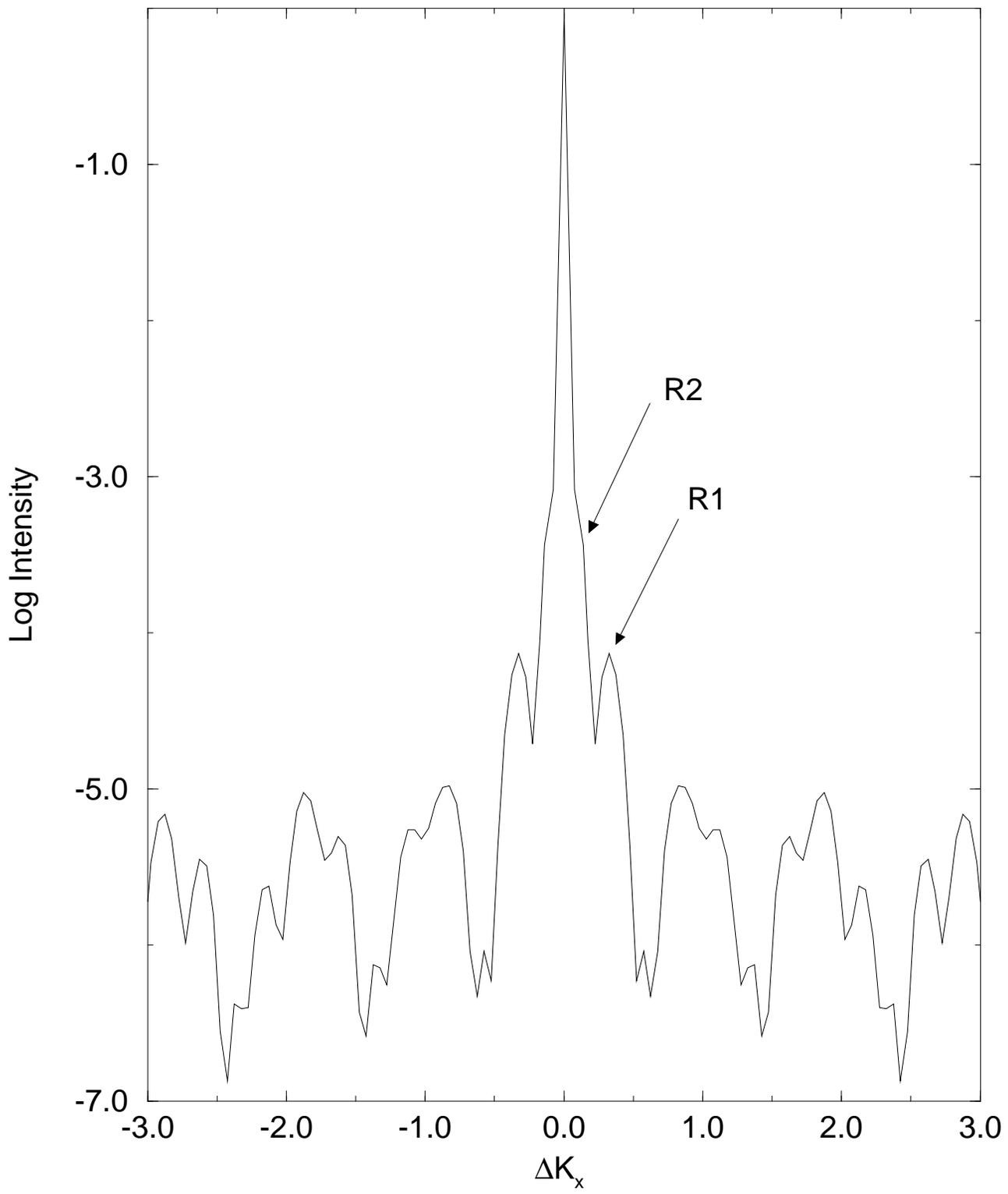
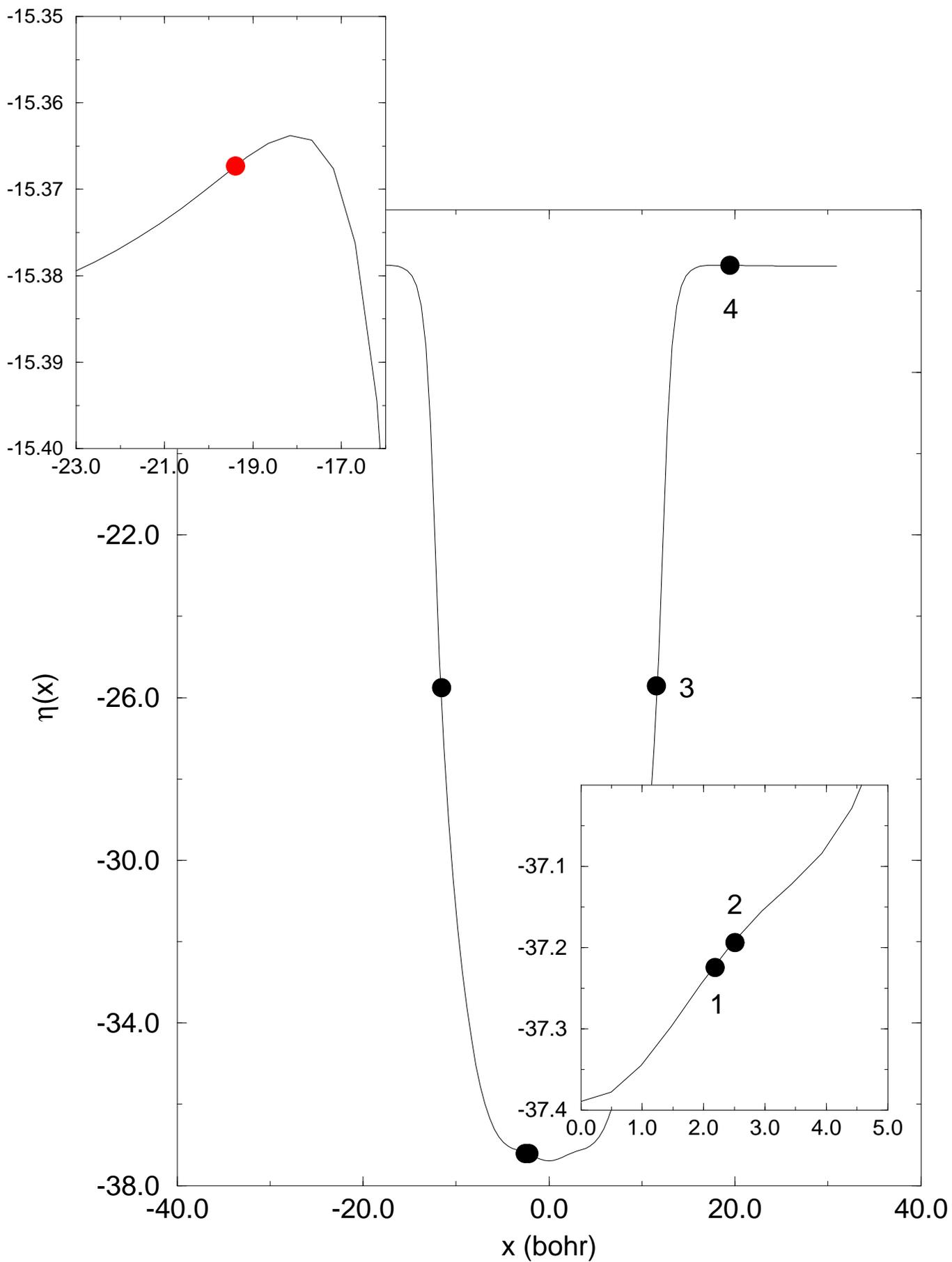
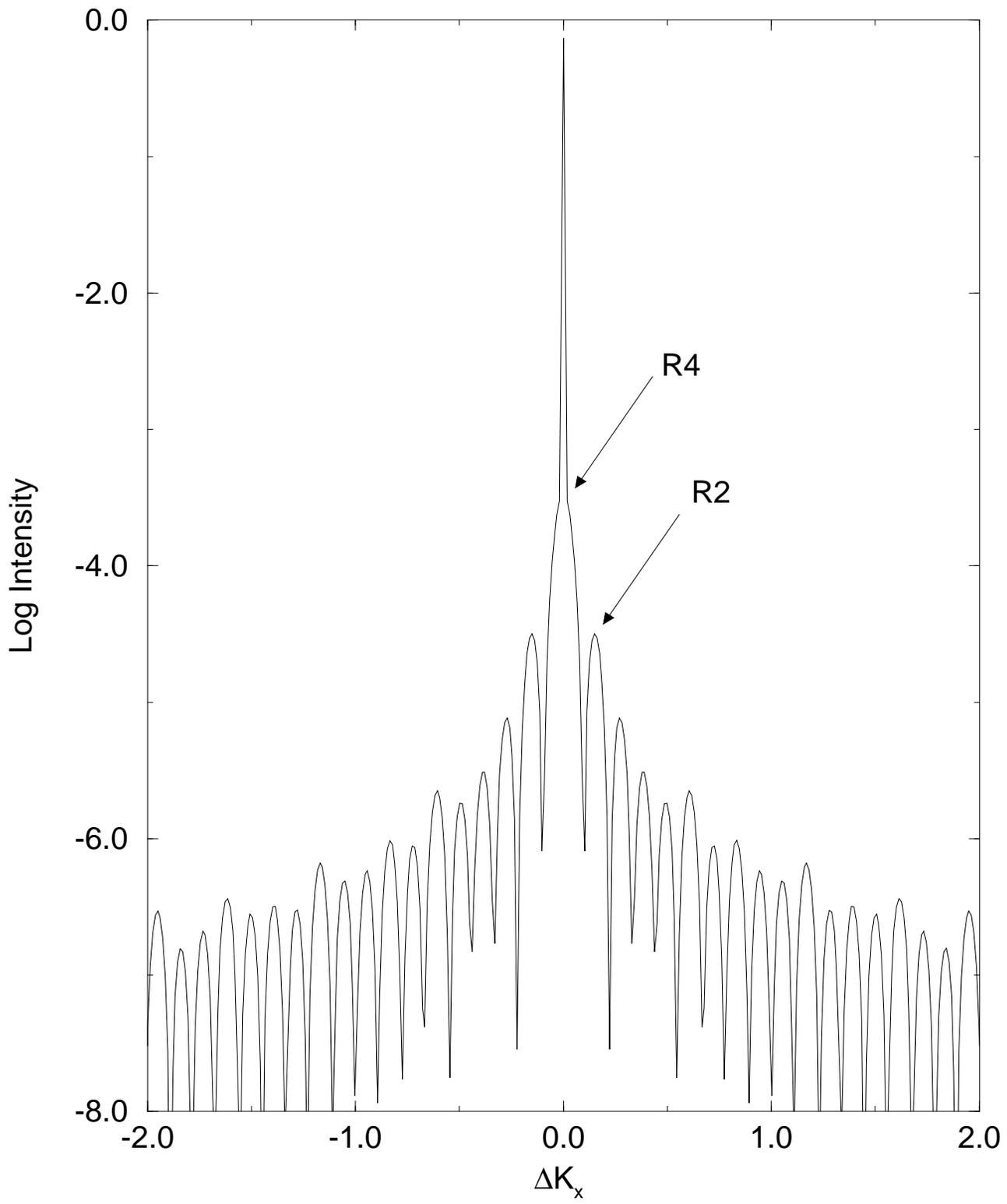
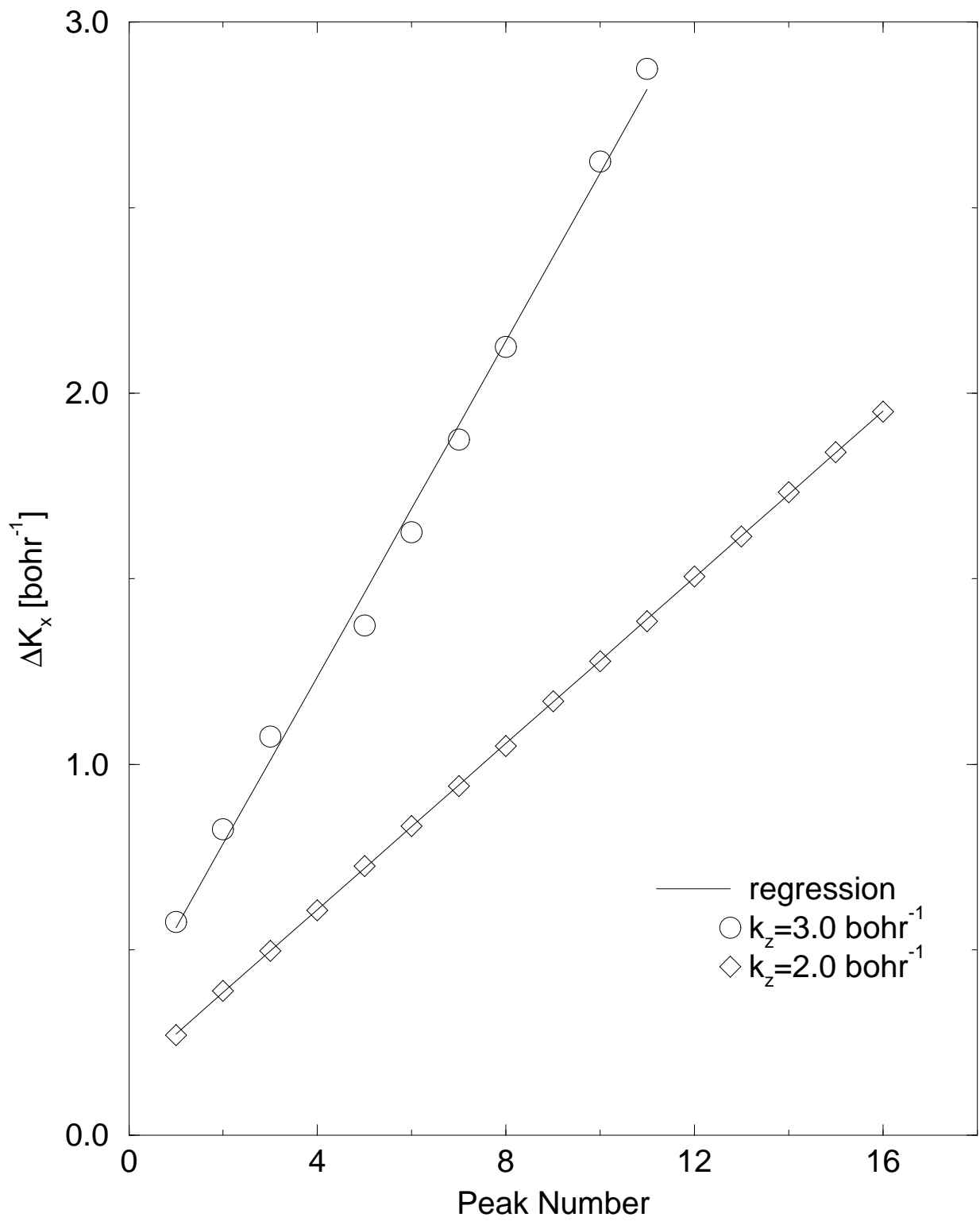
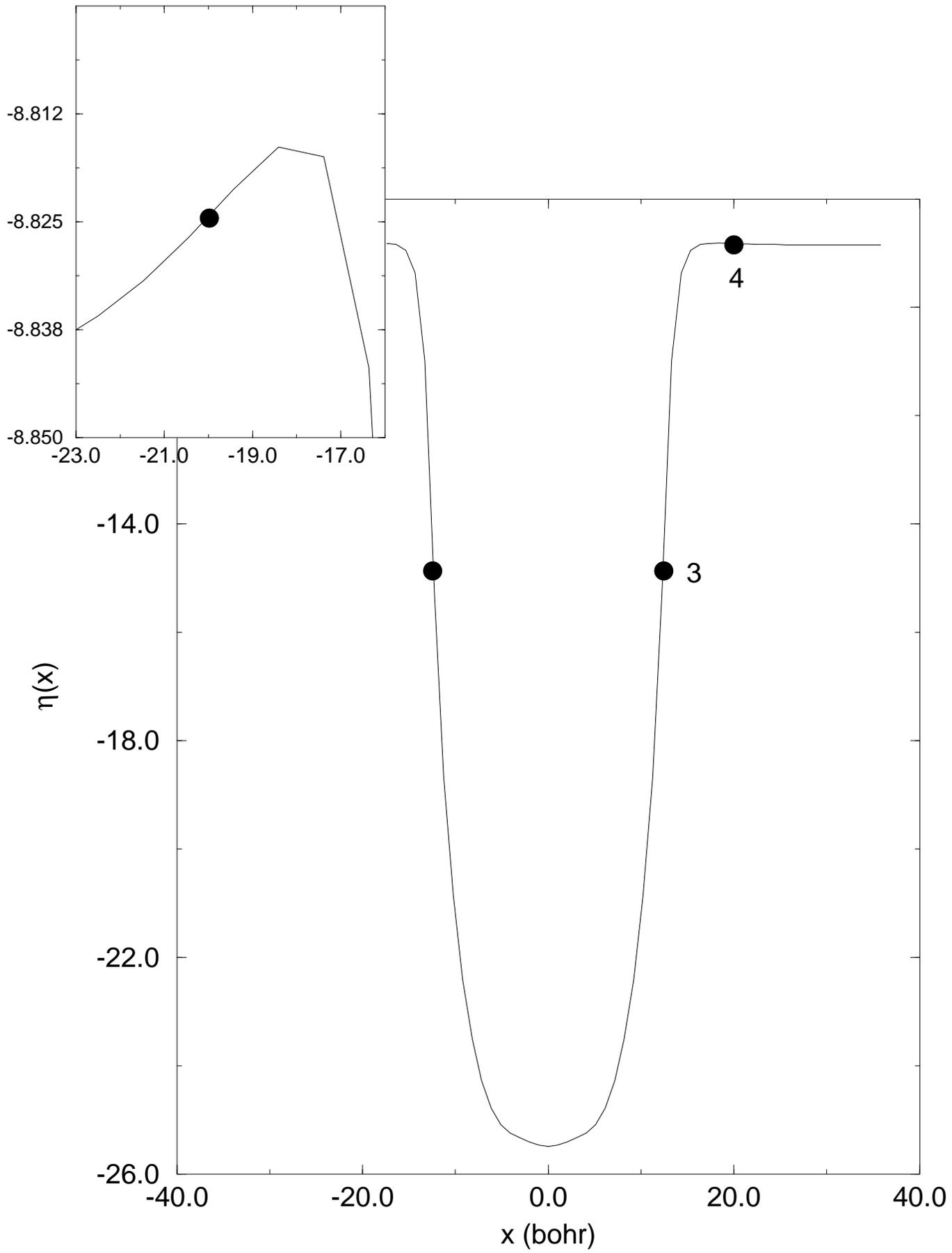
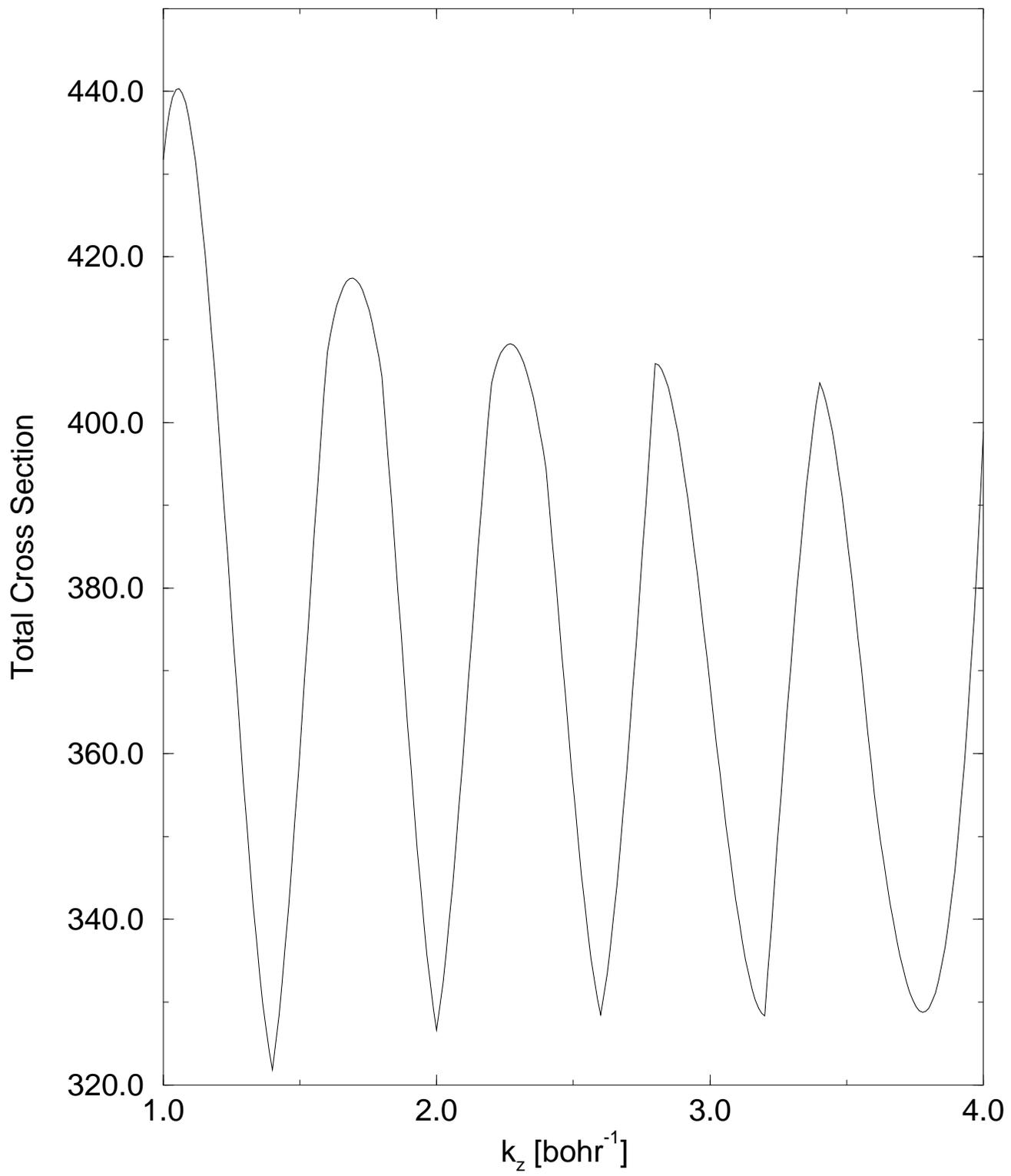
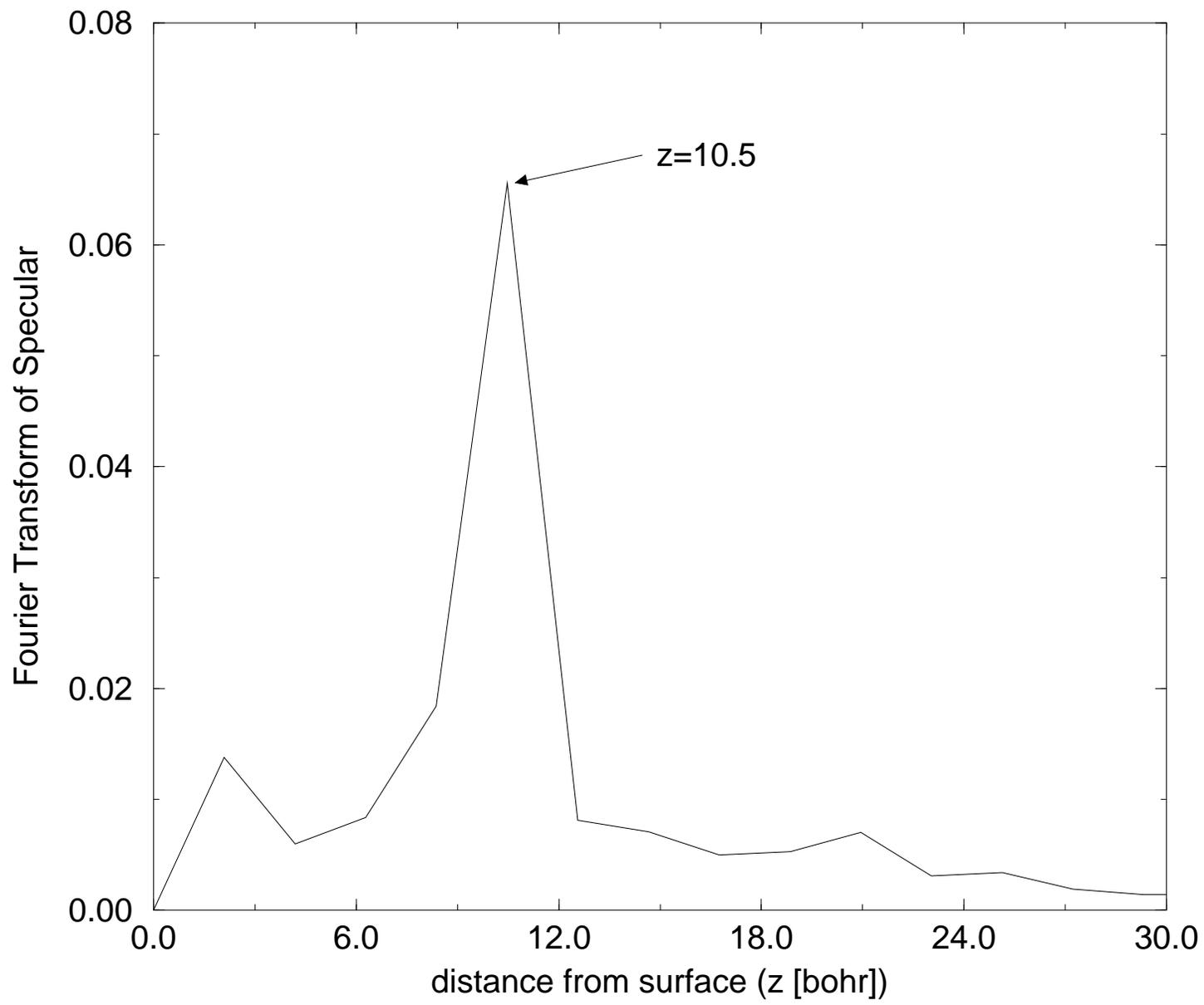
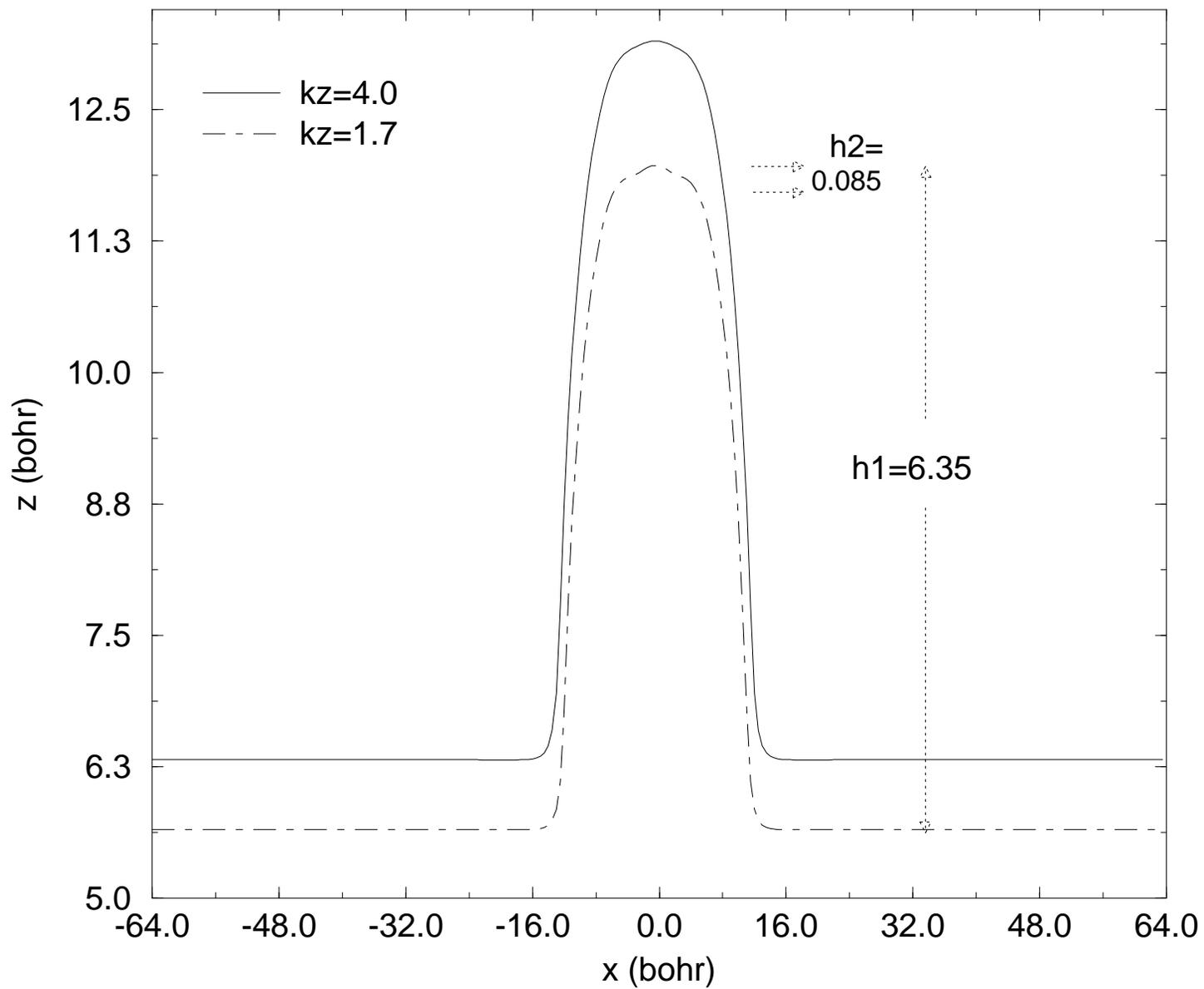
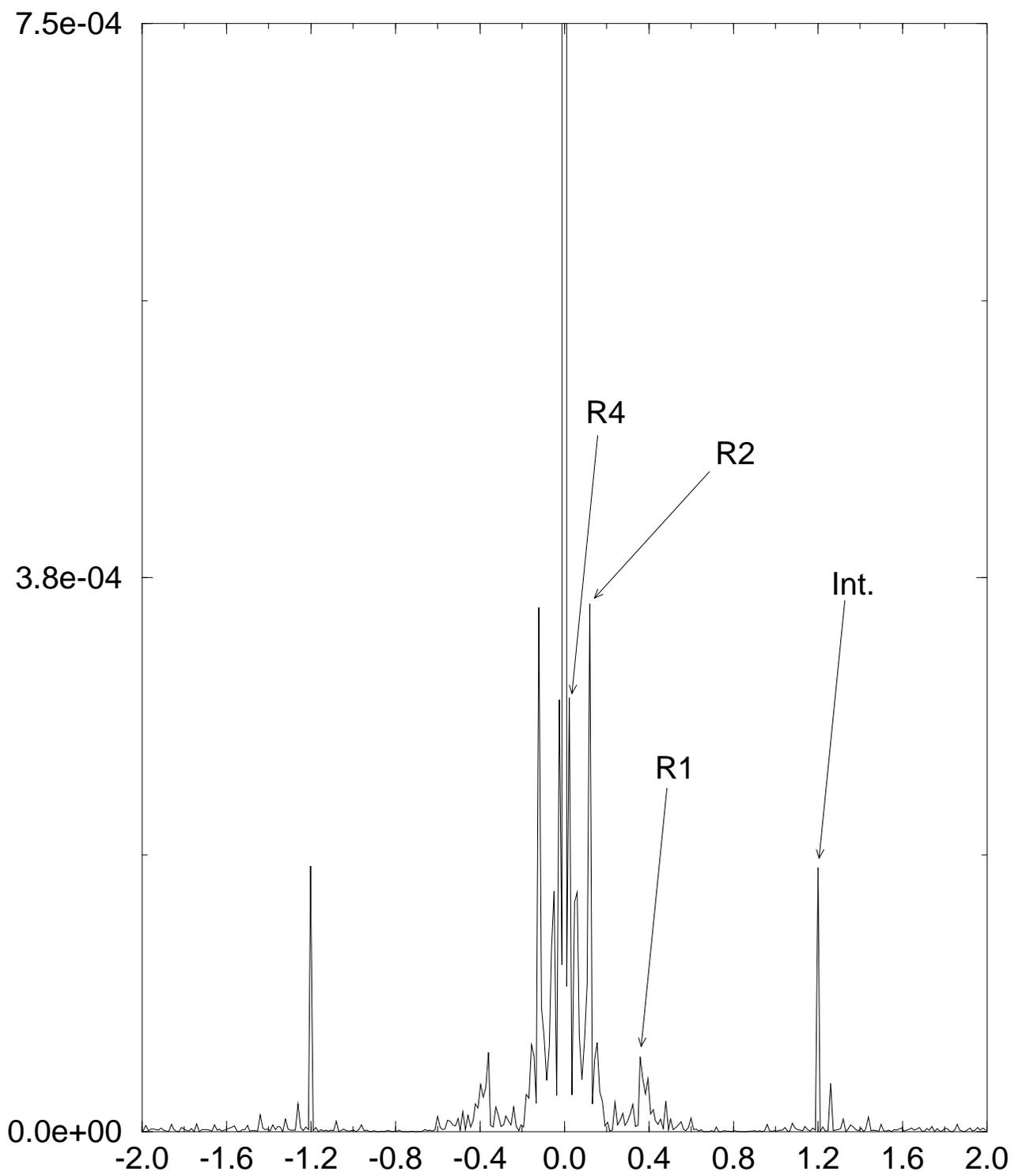
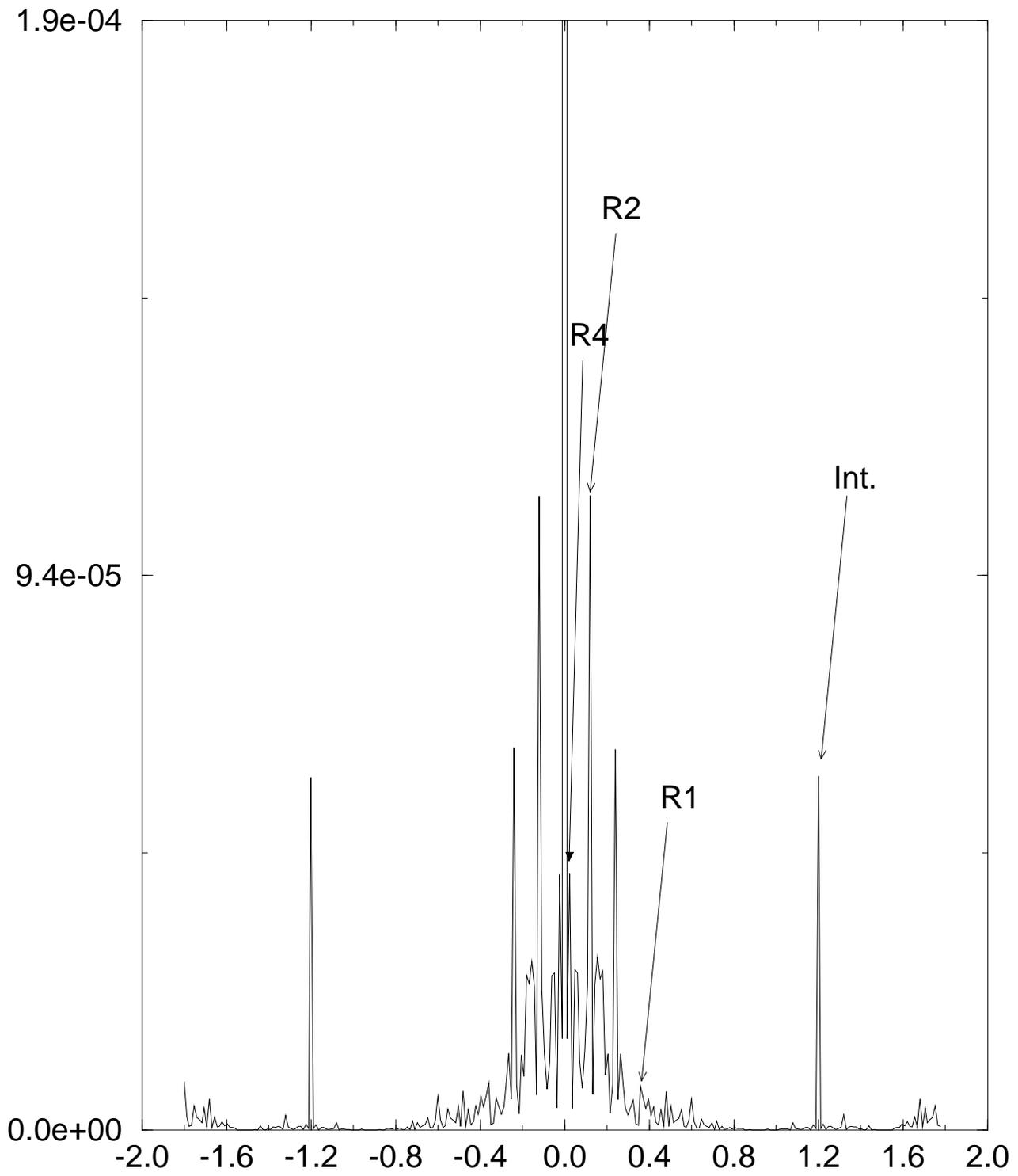
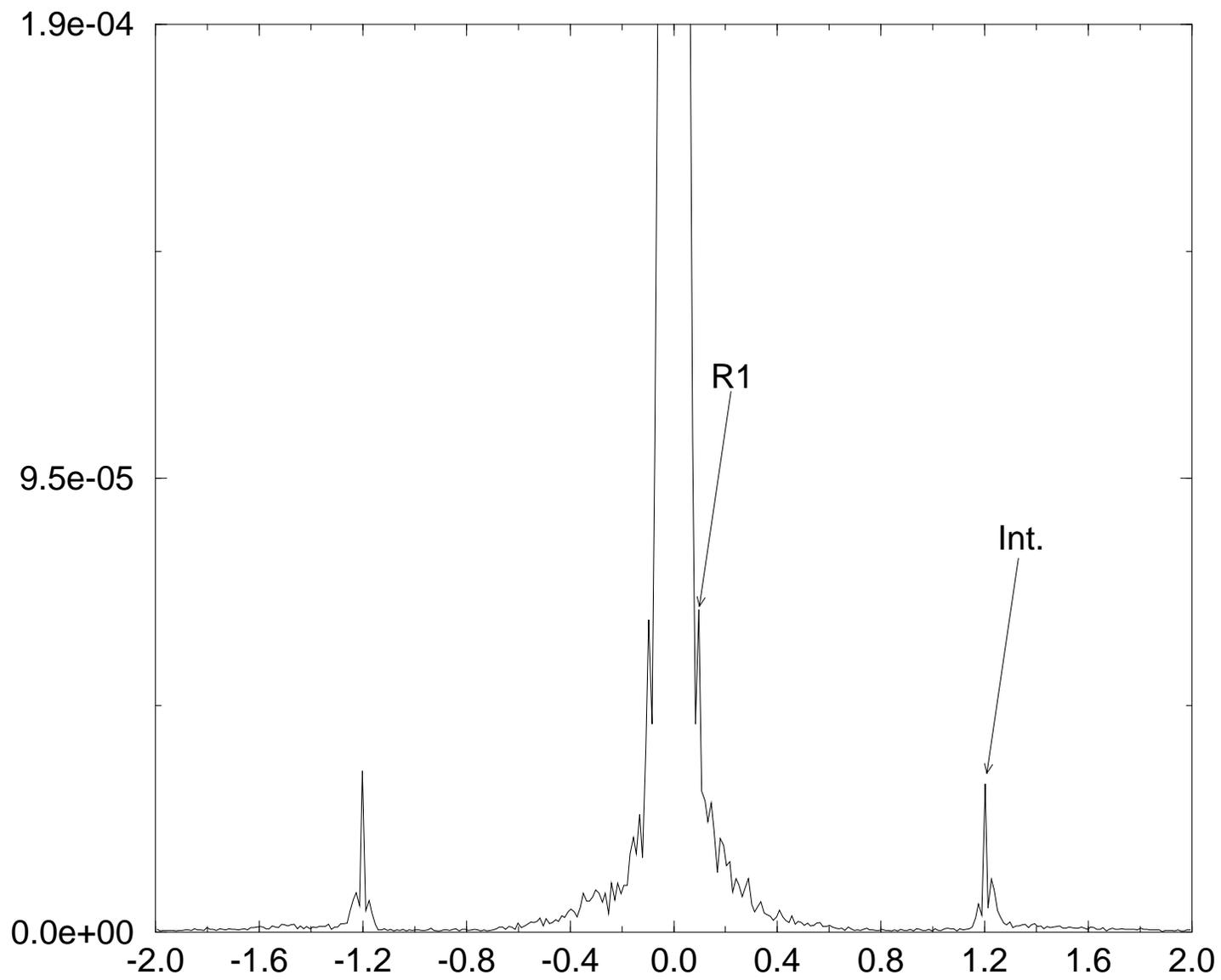
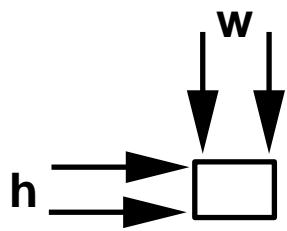
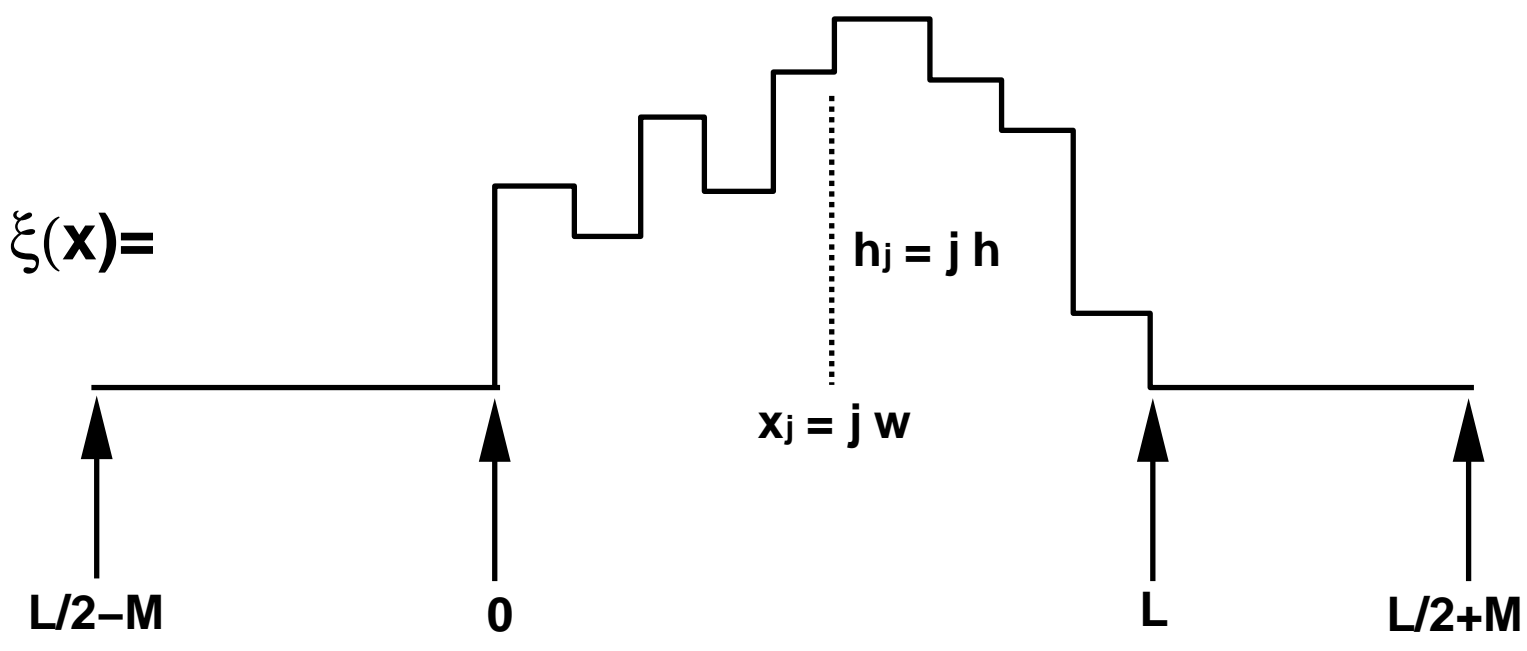
\end{document}